\documentclass[journal]{IEEEtran}
\usepackage{graphicx}
\usepackage{subcaption}
\usepackage{cite}
\usepackage{amsmath}
\usepackage{bm}
\usepackage{amsfonts}
\usepackage{array}
\usepackage{url}
\usepackage{stfloats}
\usepackage{enumitem} 
\usepackage{xcolor,soul,framed}
\usepackage{mdwmath}
\usepackage{mdwtab}
\usepackage{lineno,hyperref}

\usepackage{amssymb}
\usepackage{graphicx}
\usepackage{algorithm}
\usepackage{algpseudocode}
\usepackage{multirow}
\usepackage{booktabs}
\usepackage{cancel}

\usepackage{makecell}
\usepackage{threeparttable}

\begin{document}

\title{Perception-Aware Video Semantic Communication}

% author
\author{Yinhuan Huang,~\IEEEmembership{Student Member,~IEEE,}
        Zhijin Qin,~\IEEEmembership{Senior Member,~IEEE}

\thanks{Parts of this work have been accepted for presentation at the IEEE International Conference on Communications Workshops (ICC Workshops), 2026~\cite{pvsc-icc}. (\textit{Corresponding author: Zhijin Qin})}
\thanks{Yinhuan Huang and Zhijin Qin are with the Department of Electronic Engineering, Tsinghua University, Beijing 100084, China, and with the State Key Laboratory of Space Network and Communications, Beijing, 100084, China.
(email: huangyh24@mails.tsinghua.edu.cn; qinzhijin@tsinghua.edu.cn).}
}

\maketitle

\begin{abstract}
Ultra-high-resolution streaming and emerging immersive services are driving rapidly increasing wireless video traffic. 
However, perceptually pleasing video transmission over bandwidth-limited and latency-constrained wireless links remains challenging for conventional separated source-channel systems, which primarily target bit-level reliability and often suffer performance degradation under short-blocklength transmission.
In addition, pixel-level distortion optimization does not necessarily align with human perception, while existing learned video codecs may incur high complexity and raise deployment issues. 
This paper proposes PVSC, a perception-aware video semantic communication framework for real-time wireless video transmission. 
PVSC eliminates explicit motion-vector transmission and exploits spatio-temporal feature coding to generate compact and channel-robust symbol streams. 
It also specifies side-information formatting, reference-buffer management, and lightweight rate control, enabling stable receiver-side reconstruction and bandwidth-adaptive inference with a single model. 
Extensive experiments demonstrate that PVSC achieves superior performance across diverse datasets, resolutions, GOP configurations, and channel conditions. 
Compared with the engineered ``VTM + 5G LDPC'' baseline, PVSC saves up to about 75\% and 87\% bandwidth at comparable LPIPS and DISTS, respectively, while enabling real-time inference on a single NVIDIA RTX~4090 GPU.
\end{abstract}


\section{Introduction}
\label{sec:introduction}
\IEEEPARstart{F}{uture} video traffic is expected to grow rapidly as ultra-high-resolution live streaming becomes ubiquitous and immersive services mature~\cite{ericsson2025}. 
However, supporting streaming video services in resource-constrained wireless environments remains highly challenging. Under such conditions, limited bandwidth can severely degrade visual quality, while harsh channel conditions and limited on-device computing resources may cause playback stalls or interruptions due to decoding errors, packet losses, retransmissions, and slow codec execution.

Fundamentally, these challenges stem from the limitations of existing video communication systems based on separation-based architectures, where source coding and channel coding are designed and optimized separately.
In wireless scenarios with limited bandwidth, stringent delay constraints, and limited computing resources, systems often rely on short-blocklength channel codes~\cite{polyanskiy2010finite, durisi2016short-packets}, making it difficult to attain the optimality promised by the separation theorem~\cite{shannon1948}.
Consequently, transmission reliability becomes vulnerable to channel variations, and such systems may exhibit the cliff effect~\cite{deepwive}, causing abrupt quality degradation, reconstruction failure, or playback interruption.

\begin{figure*}[htbp]
  \centering
  \includegraphics[width=1.0\textwidth]{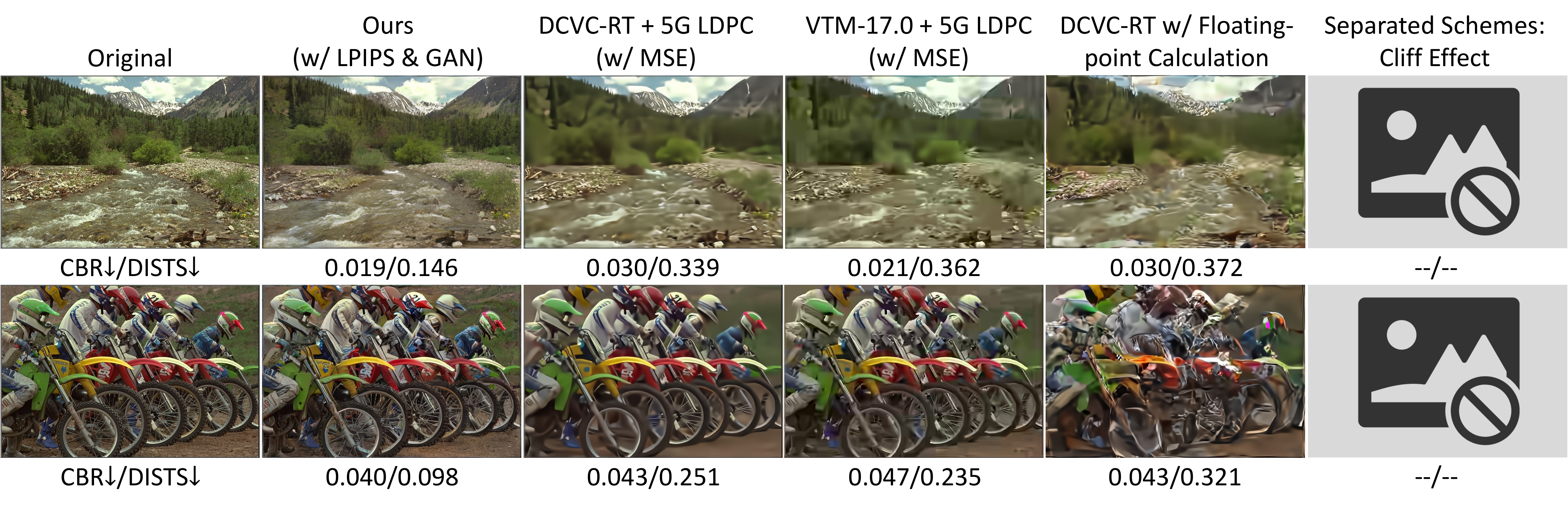}
  \caption{For bandwidth-limited transmission, pixel-level distortion optimization alone, \emph{e.g.}, MSE, does not reliably reflect perceptual quality. As shown by the ``DCVC-RT/VTM-17.0 + 5G LDPC'' scheme, LPIPS-oriented optimization achieves better perceptual quality at lower bandwidth, which is also supported by DISTS. Moreover, for DCVC-RT, cross-device floating-point discrepancies between encoding on an NVIDIA A6000 Pro GPU and decoding on an NVIDIA A800 GPU can cause severe quality degradation. In addition, separated schemes are vulnerable to the cliff effect, often leading to reconstruction failure. All transmissions shown in the figure were evaluated over an AWGN channel at SNR = 0 dB.}
  \label{motivation}
\end{figure*}

Beyond transmission reliability, perceptual quality is another major bottleneck. As shown in Fig.~\ref{motivation}, under limited bandwidth, optimizing only pixel-level distortion often fails to reflect perceptual quality and may introduce visible artifacts~\cite{mismatch, blau2018pd}.
Accordingly, conventional codecs (\emph{e.g.}, H.266/VTM~\cite{vvc}) and MSE-optimized learned video codecs (LVCs), such as DCVC~\cite{dcvc} and its variants~\cite{dcvc-dc, dcvc-fm, dcvc-rt}, become less appealing when perceptual quality is the primary objective.
A natural alternative is to directly optimize perceptual metrics such as LPIPS~\cite{lpips} and DISTS~\cite{dists}. Along this direction, several perceptual LVCs, including PLVC~\cite{plvc} and GLC-Video~\cite{glc-video}, have shown improved visual quality. However, these methods often incur higher computational complexity or latency. Moreover, as illustrated in Fig.~\ref{motivation}, LVCs are sensitive to floating-point discrepancies between the transmitter and receiver, which can degrade quality and even trigger error propagation~\cite{dcvc-rt}. Together, these issues limit the practicality of LVCs for real-time deployment on heterogeneous devices.

To address these limitations, semantic communication has recently emerged as an alternative paradigm under these constraints~\cite{10639525}. Instead of pursuing bit-level reconstruction, it transmits task-relevant semantic features and jointly optimizes source and channel processing~\cite{Task-Oriented}. This design improves efficiency and robustness under short-blocklength and computing constraints, with demonstrated gains in text~\cite{deepsc}, speech~\cite{robustsc-speech}, image~\cite{ntscc, plit, Quaddeepsc}, and point-cloud~\cite{yjk2025} transmission. 

Existing studies on video semantic communication can be broadly categorized into three representative lines: (i) causal transmission for general-scene video streaming, (ii) general-scene video transmission in non-streaming settings, and (iii) application-specific video transmission.
For (i), temporal redundancy is commonly reduced through optical-flow-based motion estimation and predictive coding~\cite{dvst, DVSC-gcwkp, wvjscc, mdvsc}, or through dynamic-static decomposition based on semantic segmentation~\cite{URLLC-iccwkp}. Semantic-rate-adaptive transmission has also been investigated, where entropy modeling generates variable-length channel symbols according to video content~\cite{dvst, wvjscc, mdvsc}. In addition, channel-adaptive schemes employ fixed-length symbols to accommodate time-varying channel conditions and alleviate the cliff effect~\cite{DVSC-gcwkp, URLLC-iccwkp}.
For (ii), unlike causal streaming schemes, non-streaming methods exploit bidirectional references and reconstruct the current frame using both past and future frames, thereby further reducing temporal redundancy~\cite{deepwive, SemCom-video}. Key-frame selection, frame interpolation, and generative reconstruction have also been introduced to improve transmission efficiency and reconstruction capability~\cite{PSD-GSC}.
For (iii), structured semantic representations and domain-specific priors are leveraged for specific applications, such as traffic surveillance~\cite{oar-vsc} and talking-head conferencing~\cite{wsc-vc, syncsc,10093128}, enabling low-rate semantic reconstruction. In parallel, task-oriented video semantic communication bypasses pixel-level reconstruction and directly supports downstream understanding tasks~\cite{videoqa-sc}.

Despite these advances, existing methods remain inadequate for perceptual-quality-oriented real-time streaming over bandwidth-limited wireless links. 
Specifically, they often (i) cannot simultaneously deliver high compression ratio and real-time streaming on consumer hardware; (ii) retain hybrid-coding designs that require multiple transmission branches, \emph{e.g.}, motion/side-information and residual-related streams, along with non-standardized side-information formats and reference-buffer management, thereby complicating synchronization and deployment; (iii) rely on bidirectional references or frame interpolation, which violate streaming causality and introduce additional latency; and (iv) lack systematic end-to-end evaluation across diverse operating points, including video resolutions, channel conditions, and group-of-pictures (GOP) configurations, with unified reporting of quality metrics, efficiency, and runtime, as well as comparisons with strong engineered separated-communication baselines.

Motivated by these gaps, we propose a perception-aware video semantic communication system (PVSC) tailored to bandwidth-limited and latency-constrained wireless links. PVSC offers a unified end-to-end pipeline and is trained to better align reconstruction with human perception. 
Extensive simulations on videos with diverse resolutions and content characteristics, under various channel conditions and GOP configurations, show that, relative to the engineered ``VTM + 5G LDPC'' baseline, PVSC achieves up to about 75\%/87\% bandwidth savings at comparable LPIPS/DISTS, respectively, while supporting real-time inference on a single NVIDIA RTX 4090 GPU. Specifically, PVSC achieves 22.7/37.0 FPS, 45.0/74.1 FPS, and 51.8/90.9 FPS at the transmitter/receiver for 1080p, 720p, and 480p videos, respectively. The main contributions of this work are summarized as follows:

\begin{itemize}
    \item We propose PVSC, a deployable end-to-end semantic video communication framework for perceptual-quality-oriented real-time streaming over bandwidth-limited wireless links. By eliminating explicit motion-vector transmission, PVSC offers a simpler and more practical solution for general-scene video streaming.
    
    \item We develop a spatio-temporal feature coding strategy that exploits diverse spatial and temporal contexts to efficiently generate compact symbol streams while improving robustness against channel impairments.
    
    \item We design an inference pipeline with a specified side-information format and standardized reference-buffer management, ensuring consistent reference alignment and stable receiver-side reconstruction. Moreover, we introduce a lightweight rate-control mechanism for bandwidth-adaptive inference with a single model, reducing both training and deployment costs.
\end{itemize}

The remainder of this paper is organized as follows. 
Section~\ref{sec:system_model} introduces the system model. 
Section~\ref{sec:architecture} details the architecture and implementation. 
Section~\ref{sec:experiments} reports the experimental results and analysis. 
Section~\ref{sec:conclusion} concludes the paper.

\textit{Notation:} Scalars are denoted by lowercase letters (\emph{e.g.}, $x$), vectors by bold lowercase letters (\emph{e.g.}, $\mathbf{x}$), and matrices by bold uppercase letters (\emph{e.g.}, $\mathbf{X}$). The sets $\mathbb{C}^{m \times n}$ and $\mathbb{R}^{m \times n}$ denote the spaces of $m \times n$ complex-valued and real-valued matrices, respectively. The probability density function of a continuous random variable $x$ is written as $p_x$, and the probability mass function of a discrete random variable $\bar{x}$ is written as $P_{\bar{x}}$. The operator $\mathbb{E}[\cdot]$ denotes expectation, $||\cdot||_2$ denotes the euclidean norm, and $\odot$ denotes element-wise multiplication. Neural network parameters are collected in $\boldsymbol{\theta}$, with subscripts indicating the corresponding module (\emph{e.g.}, $\boldsymbol{\theta}_{\mathcal{G}}$). $\mathcal{N}(x|\mu, \sigma^2)$ denotes a gaussian distribution with mean $\mu$ and variance $\sigma^2$. Superscripts $tx$ and $rx$ distinguish decoded quantities obtained at the transmitter and receiver, respectively, \emph{e.g.}, $\mathbf{F}_t^{tx}$ and $\mathbf{F}_t^{rx}$.

\section{System Model and Problem Formulation}
\label{sec:system_model}
This section describes the system model of PVSC, including the problem formulation, framework and loss functions.

\begin{figure*}[htbp]
  \centering
  \includegraphics[width=1.0\textwidth]{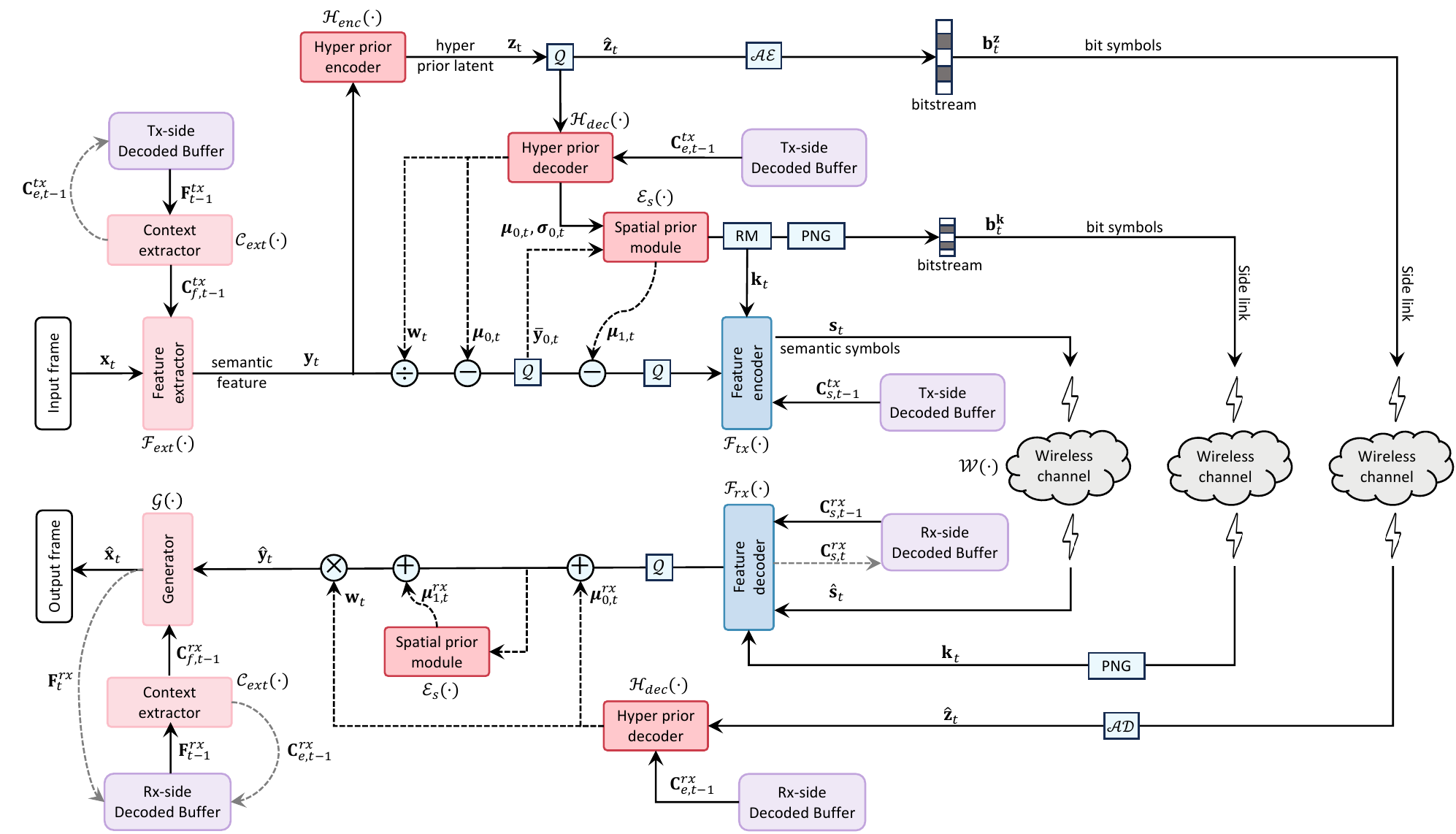}
  \caption{System model of the proposed PVSC. $\mathcal{AE}$, $\mathcal{AD}$, $\mathcal{Q}$, PNG, and RM denote arithmetic encoding, arithmetic decoding, quantization, portable network graphics coding, and rate matching, respectively. PVSC models implicit spatio-temporal dependencies, where $\mathbf{F}_{t-1}^{tx/rx}$ is used to generate $\mathbf{C}_{e,t-1}^{tx/rx}$ and $\mathbf{C}_{f,t-1}^{tx/rx}$. $\mathbf{C}_{e,t-1}^{tx/rx}$, $\mathbf{C}_{f,t-1}^{tx/rx}$, and $\mathbf{C}_{s,t-1}^{tx/rx}$ denote the temporal contexts used at time $t$ for entropy estimation, feature-domain processing, and feature-symbol mapping, respectively.}
  \label{system_model}
\end{figure*}

\subsection{Wireless Video Transmission Scenario}
We consider a point-to-point video transmission system over a wireless link between two endpoints, such as an edge server, a base station, or a mobile device. 
The sending and reconstruction endpoints are referred to as the transmitter and receiver, respectively. 
At the transmitter, the input video is denoted by $\{\mathbf{x}_t\}_{t=1}^T$, where $\mathbf{x}_t \in \mathbb{R}^{H \times W \times C}$ is the $t$-th frame, and $H$, $W$, and $C$ denote the height, width, and number of color channels, respectively. 
In a low-latency video communication scenario, $T$ consecutive frames are grouped into a GOP. 
The first frame is intra-coded as an I-frame, whereas the remaining frames are encoded as P-frames with reference to the preceding frame. 
The transmitter encodes $\{\mathbf{x}_t\}_{t=1}^T$ into complex symbol sequences $\{\mathbf{s}_t\}_{t=1}^T$ with $\mathbf{s}_t \in \mathbb{C}^{l_t}$, where $l_t$ denotes the number of channel uses at time step $t$. 
The receiver reconstructs $\{\hat{\mathbf{x}}_t\}_{t=1}^T$ from the received symbol sequences $\{\hat{\mathbf{s}}_t\}_{t=1}^T$, targeting perceptually high-quality video recovery.

The wireless channel is modeled by the transfer function:
\begin{equation}
  \label{eq:channel}
  \mathcal{W}(\mathbf{s}_t)=\mathbf{h}_t \odot \mathbf{s}_t + \mathbf{n}_t,
\end{equation}
where $\mathbf{h}_t\in \mathbb{C}^{l_t}$ is the complex channel coefficient and $\mathbf{n}_t\!\sim\!\mathcal{CN}(\mathbf{0},\sigma_{noise}^2\mathbf{I}_{l_t})$ is additive white gaussian noise (AWGN). 
The transmitted symbols satisfy the power constraint $\mathbb{E}_{\mathbf{s}_t}\!\left[\|\mathbf{s}_t\|_2^2\right]\leq P$. Accordingly, the average signal-to-noise ratio (SNR) is defined as:
\begin{equation}
  \label{eq:snr}
  \begin{aligned}
     \text{SNR} = 10 \log_{10} \left( \frac{\mathbb{E}_{\mathbf{h}_t}\left[\lVert \mathbf{h}_t\rVert_2^2\right]P}{\sigma_{noise}^2} \right) 
      \geq \\
      10 \log_{10} \left( \frac{\mathbb{E}_{\mathbf{h}_t}\left[\lVert \mathbf{h}_t\rVert_2^2\right]\mathbb{E}_{\mathbf{s}_t} \left[||\mathbf{s}_t||_2^2\right]}{\mathbb{E}_{\mathbf{n}_t} \left[||\mathbf{n}_t||_2^2\right]} \right),
  \end{aligned}
\end{equation}
where $\sigma_{noise}^2$ denotes the average noise power. For the AWGN case, $\mathbf{h}_t$ is time-invariant. For rayleigh fading, we adopt a block-fading model where the channel remains constant within each coherence interval of $L_{co}$ channel uses and changes independently across intervals. Specifically, for the $t$-th interval, $h_{t,i}\overset{\text{i.i.d.}}{\sim}\mathcal{CN}(0,1)$ and $h_{t,i}=h_{t,1}$ for $i=1,\ldots,L_{co}$. The estimated channel state information (CSI) is modeled as
$\hat{\mathbf{h}}_t=\mathbf{h}_t+\mathbf{e}_t$, where $\mathbf{e}_t\sim\mathcal{CN}(0,\sigma_e^2\mathbf{I}_l)$. The CSI accuracy is measured by the normalized mean-squared error (NMSE)~\cite{bjornson2017massive}:
\begin{equation}
  \label{nmse}
  \mathrm{NMSE}_{\mathrm{dB}}=10\log_{10}(\frac{\mathbb{E}\left[\lVert \mathbf{e}_t\rVert_2^2\right]}{\mathbb{E}\left[\lVert \mathbf{h}_t\rVert_2^2\right]}).
\end{equation}
Based on $\hat{\mathbf{h}}_t$, the receiver applies zero-forcing (ZF) equalization as:
\begin{equation}
  \label{eq:zfrx}
    \hat{\mathbf{s}}_t \leftarrow \frac{\hat{\mathbf{h}}_t^*}{|\hat{\mathbf{h}}_t|^2} \odot \mathcal{W}(\mathbf{s}_t),
\end{equation}
where $\hat{\mathbf{h}}^*_t$ is the complex conjugate of $\hat{\mathbf{h}}_t$. 

We denote the source bandwidth by $n=T\times H \times W \times C$ and the channel bandwidth by $\sum_{t=1}^T{l_t}$. The channel bandwidth ratio (CBR)~\cite{deepjscc} is defined as:\footnote{In simulations, the CBR also accounts for side information transmitted in addition to complex symbol sequences $\{\mathbf{s}_t\}_{t=1}^T$.} 
\begin{equation}
  \label{eq:zfimp}
  \text{CBR}\triangleq\sum_{t=1}^T{\frac{l_t}{n}}=\sum_{t=1}^T{\frac{l_t}{T\times H \times W \times C}},
\end{equation}
which measures bandwidth efficiency as the number of channel uses per source sample.

\begin{figure*}[htbp]
  \centering
  \includegraphics[width=1.0\textwidth]{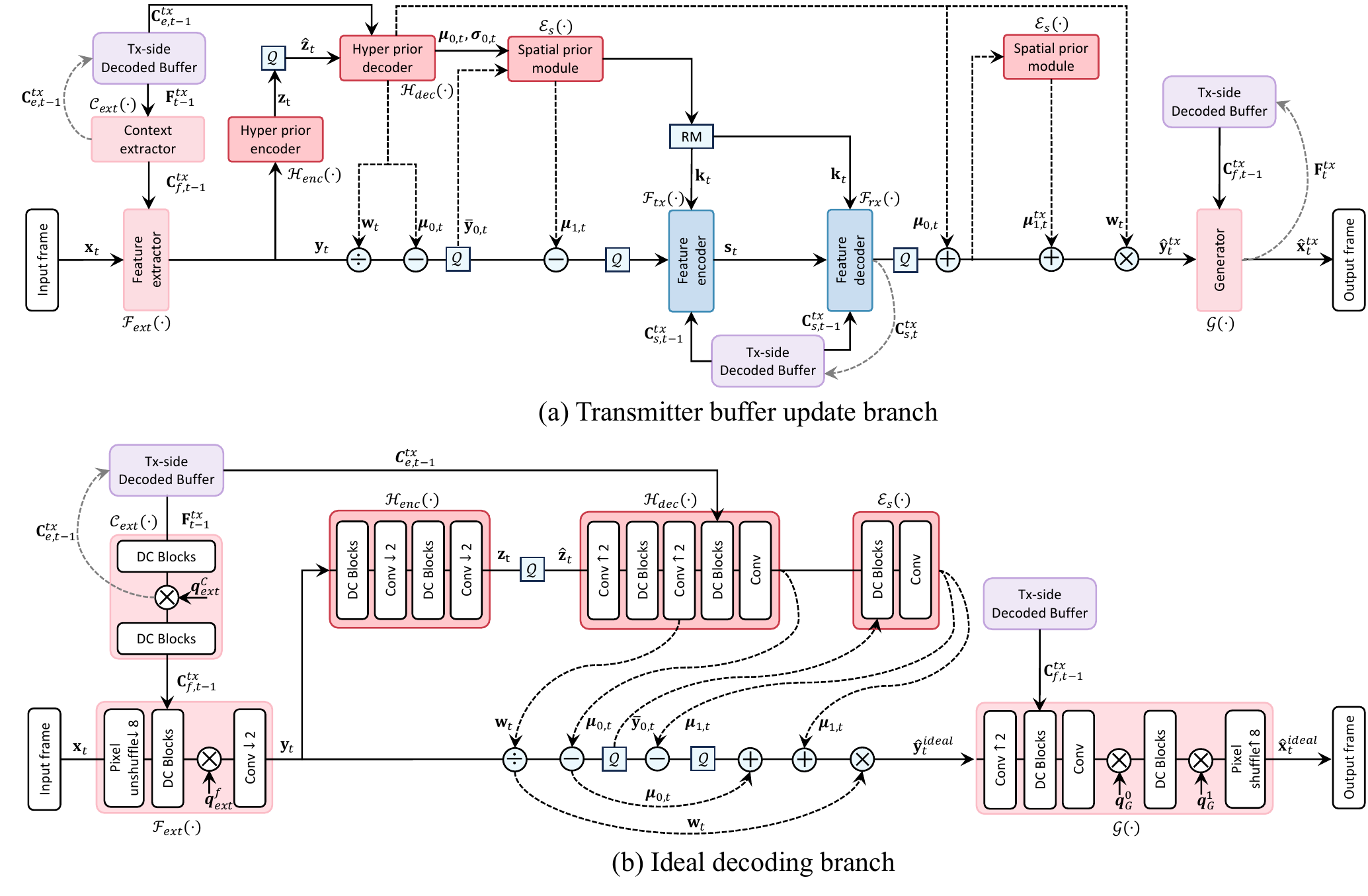}
  \caption{(a) Transmitter-side buffer update branch for local buffer updating and temporal-context generation $(\mathbf{C}_{s,t}^{tx}, \mathbf{F}_{t}^{tx})$. (b) Ideal decoding branch. All modules use DC blocks with learnable vectors $(\mathbf{q}_{ext}^{C}, \mathbf{q}_{ext}^{f}, \{\mathbf{q}_{G}^{i}\}_{i=0}^{1})$ for multi-rate control.}
  \label{Branch}
\end{figure*}

\subsection{The Framework of PVSC}
\subsubsection{Overview and Design Principles}
Fig.~\ref{system_model} provides an overview of PVSC. For each input frame $\mathbf{x}_t$, PVSC generates two types of streams: a compact complex symbol sequence $\mathbf{s}_t$ transmitted over the bandwidth-limited wireless channel,
and two side-information bitstreams, namely the hyper-latent bitstream $\mathbf{b}_t^{\mathbf{z}}$ and the rate-map bitstream $\mathbf{b}_t^{\mathbf{k}}$, delivered through a side link. The receiver jointly decodes these streams to
reconstruct $\hat{\mathbf{x}}_t$ and update the reference buffer for subsequent frames. 
This workflow is guided by three design principles. 
(i) Deployable causal semantic streaming: for real-time streaming, PVSC replaces costly explicit motion signaling with implicit temporal modeling based on decoded-buffer features from the previous frame. 
(ii) Spatio-temporal feature coding: PVSC estimates spatially nonuniform feature redundancy using hyperprior and spatio-temporal priors, and performs reference-conditioned feature-to-symbol mapping to suppress redundancy and generate compact, channel-robust symbols for reliable transmission.
(iii) Deployment-oriented inference and rate adaptation: to avoid transmitter-receiver reference mismatch, PVSC standardizes side information and reference-buffer management, introduces a local transmitter update branch, and supports single-model bandwidth adaptation via lightweight rate control.

\subsubsection{Transmitter-Side Processing}
The transmitter-side processing pipeline comprises three main stages: temporal-context-aided feature extraction, probabilistic feature quantization, and rate-aware symbol mapping.

PVSC employs separate buffers at the transmitter and the receiver. Each buffer caches intermediate features from last time step $t-1$ and uses them as temporal context for coding the current frame. This design enables temporally aware feature representation without explicitly transmitting motion information. 
Given the intermediate feature $\mathbf{F}_{t-1}^{tx}$ produced by the generator $\mathcal{G}(\cdot)$ at time step $t-1$, the context extractor $\mathcal{C}_{ext}(\cdot)$ generates two temporal reference tensors:
\begin{equation}
  \label{eq:ctx_ext}
  (\mathbf{C}_{e,t-1}^{tx}, \mathbf{C}_{f,t-1}^{tx}) = \mathcal{C}_{ext}(\mathbf{F}_{t-1}^{tx}; \boldsymbol{\theta}_{\mathcal{C}_{ext}}).
\end{equation}
Conditioned on $\mathbf{C}_{f,t-1}^{tx}$, the feature extractor $\mathcal{F}_{ext}(\cdot)$ then maps the current video frame $\mathbf{x}_t$ to a semantic feature $\mathbf{y}_t$:
\begin{equation}
  \label{eq:feauture_ext}
  \mathbf{y}_t = \mathcal{F}_{ext}(\mathbf{x}_t \mid \mathbf{C}_{f,t-1}^{tx}; \boldsymbol{\theta}_{\mathcal{F}_{ext}}).
\end{equation}

For probabilistic feature quantization, $\mathbf{y}_t$ is modeled as a gaussian random vector with spatially varying parameters, following~\cite{balle2016end,dcvc,dvst}. 
Its distribution is inferred from three types of causal information: a hyperprior that captures global feature statistics, a spatial reference that exploits dependencies among previously processed feature blocks, and temporal contexts extracted from the previous decoded feature.
The probability model therefore consists of a hyperprior module, parameterized by an encoder $\mathcal{H}_{enc}(\cdot)$ and a decoder $\mathcal{H}_{dec}(\cdot)$, together with a spatial prior module $\mathcal{E}_s(\cdot)$.

The $\mathcal{H}_{enc}(\cdot)$ first produces the hyperprior latent $\mathbf{z}_t$, which is then quantized to $\hat{\mathbf{z}}_t = \mathcal{Q}(\mathbf{z}_t)$ and entropy coded using a learned factorized prior: 
\begin{equation}
  \label{eq:h_enc}
  \mathbf{b}_t^\mathbf{z} = \mathcal{AE}\big(\mathcal{Q}\big(\mathcal{H}_{enc}(\mathbf{y}_t; \boldsymbol{\theta}_{\mathcal{H}_{enc}})\big)\big),
\end{equation}
where $\mathcal{AE}(\cdot)$ and $\mathcal{AD}(\cdot)$ denote arithmetic encoding and arithmetic decoding, respectively, 
with:
\begin{equation}
  \label{eq:z_dec}
  \hat{\mathbf{z}}_t = \mathcal{AD}(\mathbf{b}_t^\mathbf{z}).
\end{equation}
The resulting bitstream $\mathbf{b}_t^{\mathbf{z}}$ is transmitted to the receiver through a side link using low-density parity-check (LDPC) coding and quadrature amplitude modulation (QAM). 
The $\mathcal{H}_{dec}(\cdot)$ then combines $\hat{\mathbf{z}}_t$ with the temporal context $\mathbf{C}_{e,t-1}^{tx}$ to generate the initial distribution parameters:
\begin{equation}
  \label{eq:h_dec}
  (\mathbf{w}_t, \boldsymbol{\mu}_{0,t}, \boldsymbol{\sigma}_{0,t}) = \mathcal{H}_{dec}(\hat{\mathbf{z}}_t, \mathbf{C}_{e,t-1}^{tx}; \boldsymbol{\theta}_{\mathcal{H}_{dec}}),
\end{equation}
where $\mathbf{w}_t$, $\boldsymbol{\mu}_{0,t}$, and $\boldsymbol{\sigma}_{0,t}$ denote the quantization step, mean, and variance, respectively. 

Based on these initial parameters, 
$\mathbf{y}_t$ and $\mathbf{w}_t$ are spatially partitioned into $N$ blocks, denoted by $\{\mathbf{y}_{i,t}\}_{i=0}^{N-1}$ and $\{\mathbf{w}_{i,t}\}_{i=0}^{N-1}$. 
Let $\mathcal{I}$ denote the set of block indices. For the first block, the initial parameters are directly used. For each subsequent block $i\in\mathcal{I}\setminus\{0\}$, the $\mathcal{E}_s(\cdot)$ refines the block-wise distribution parameters using previously processed blocks:
\begin{equation}
  \label{eq:Es}
  (\boldsymbol{\mu}_{i,t}, \boldsymbol{\sigma}_{i,t})
  = \mathcal{E}_s(\mathbf{w}_t, \boldsymbol{\mu}_{0,t}, \boldsymbol{\sigma}_{0,t},
  \mathbf{y}_{<i,t}; \boldsymbol{\theta}_{\mathcal{E}_s}).
\end{equation}
The corresponding quantized representation is obtained as:
\begin{equation}
  \label{eq:quad}
  \bar{\mathbf{y}}_{i,t}
  = \mathcal{Q}\left( \frac{\mathbf{y}_{i,t}}{\mathbf{w}_{i,t}}
  - \boldsymbol{\mu}_{i,t} \right),
  \quad \forall i\in\mathcal{I}.
\end{equation}
Accordingly, the conditional probability of $\bar{\mathbf{y}}_t$ is modeled as:
\begin{equation}
  \label{eq:prob_bpp_1}
  \begin{aligned}
    p_{\bar{\mathbf{y}}_t}\!\left(
      \bar{\mathbf{y}}_t \mid 
      \hat{\mathbf{z}}_t,\mathbf{C}_{e,t-1}^{tx},\mathbf{C}_{f,t-1}^{tx}
    \right)
    = \prod_i p_{\bar{y}_{i,t}} (0,\sigma_{i,t}) \\
    = \prod_i \bigg(
         \mathcal{N}(0,\sigma_{i,t}^2)
         * \mathcal{U}(-\tfrac{1}{2},\tfrac{1}{2})
       \bigg)(\bar{y}_{i,t}).
  \end{aligned}
\end{equation}
Here, $\bar{\mathbf{y}}_t \in \mathbb{R}^{H_y \times W_y \times C_y}$ contains $H_y \times W_y$ spatial units $\bar{\mathbf{y}}_t^{(i,j)}$. Each unit corresponds to a channel symbol $\mathbf{s}_t^{(i,j)} \in \mathbb{C}^{l_t^{(i,j)}}$ with $l_t^{(i,j)} = \lceil k_t^{(i,j)} / 2 \rceil$. The symbol-length factor $k_t^{(i,j)}$ is calculated from the accumulated entropy of $\bar{\mathbf{y}}_t^{(i,j)}$ as:
\begin{equation}
  \label{eq:kij}
  \begin{aligned}
  k_t^{(i,j)} 
  = \mathcal{Q}\big( \sum_{k=1}^{C_y} -\eta \log_{2} P_{\bar{\mathbf{y}}_t} (\bar{y}_t^{(i,j,k)} \mid \hat{\mathbf{z}}_t, \mathbf{C}_{e,t-1}^{tx}, \mathbf{C}_{f,t-1}^{tx}) \big) \\
  = \mathcal{Q}\big( \sum_{k=1}^{C_y} -\eta \log_{2} \big( \Phi\!\left(\frac{\bar{y}_t^{(i,j,k)} + \tfrac{1}{2}}{\sigma_t^{(i,j,k)}}\right)
    - \Phi\!\left(\frac{\bar{y}_t^{(i,j,k)} - \tfrac{1}{2}}{\sigma_t^{(i,j,k)}}\right) \big) \big),
  \end{aligned}
\end{equation}
where $\eta$ controls the spectral efficiency and $\Phi(\cdot)$ denotes the cumulative distribution function of the standard normal distribution. To reduce search complexity, $k_t^{(i,j)}$ is quantized to the nearest value in a predefined discrete set, and its index is stored in the rate map $\mathbf{k}_t \in \mathbb{R}^{H_y \times W_y}$. This process is referred to as rate matching (RM) in Fig.~\ref{system_model}. We use lossless portable network graphics (PNG)~\cite{png-guide} to compress $\mathbf{k}_t$ into a bitstream $\mathbf{b}_t^{\mathbf{k}}$, 
which is transmitted in the same manner as $\mathbf{b}_t^{\mathbf{z}}$.

Finally, the feature encoder $\mathcal{F}_{tx}(\cdot)$ uses $\mathbf{k}_t$ to select rate embeddings and reweight the features. 
Conditioned on the previous-step decoded state $\mathbf{C}_{s,t-1}^{tx}$, it maps $\bar{\mathbf{y}}_t$ into the complex symbol sequence $\mathbf{s}_t$:
\begin{equation}
  \label{eq:ftx}
 \mathbf{s}_t = \mathcal{F}_{tx}(\bar{\mathbf{y}}_t,  \mathbf{k}_t \mid \mathbf{C}_{s,t-1}^{tx}; \boldsymbol{\theta}_{\mathcal{F}_{tx}}).
\end{equation}

\subsubsection{Receiver-Side Reconstruction} 
Given the equalized symbols and side information, the receiver first recovers the quantized feature and then performs probabilistic feature dequantization. 
Specifically, using $\mathbf{k}_t$ and the previous-step decoded state $\mathbf{C}_{s,t-1}^{rx}$, the feature decoder $\mathcal{F}_{rx}(\cdot)$ recovers $\bar{\mathbf{y}}_t^{rx}$ from $\hat{\mathbf{s}}_t$ 
and updates the state $\mathbf{C}_{s,t}^{rx}$ for the next time step:
\begin{equation}
  \label{eq:frx1}
 (\bar{\mathbf{y}}_t^{rx}, \mathbf{C}_{s,t}^{rx}) = \mathcal{F}_{rx}(\hat{\mathbf{s}}_t,  \mathbf{k}_t \mid \mathbf{C}_{s,t-1}^{rx}; \boldsymbol{\theta}_{\mathcal{F}_{rx}}).
\end{equation}

For temporal conditioning, the context extractor $\mathcal{C}_{ext}(\cdot)$ takes the previous intermediate feature $\mathbf{F}_{t-1}^{rx}$ and produces:
\begin{equation}
  \label{eq:ctx_ext_rx}
  (\mathbf{C}_{e,t-1}^{rx}, \mathbf{C}_{f,t-1}^{rx}) = \mathcal{C}_{ext}(\mathbf{F}_{t-1}^{rx}; \boldsymbol{\theta}_{\mathcal{C}_{ext}}).
\end{equation}

The hyperprior decoder $\mathcal{H}_{dec}(\cdot)$ estimates the initial dequantization parameters from $\hat{\mathbf{z}}_t$ and $\mathbf{C}_{e,t-1}^{rx}$ as:
\begin{equation}
  \label{eq:h_dec_rx}
  (\mathbf{w}_t, \boldsymbol{\mu}_{0,t}^{rx}, \boldsymbol{\sigma}_{0,t}^{rx}) = \mathcal{H}_{dec}(\hat{\mathbf{z}}_t, \mathbf{C}_{e,t-1}^{rx}; \boldsymbol{\theta}_{\mathcal{H}_{dec}}).
\end{equation}

With the same spatial partition, for each block $i\in\mathcal{I}\setminus\{0\}$, the spatial prior module $\mathcal{E}_s(\cdot)$ predicts the block-wise distribution parameters from the previously recovered blocks:
\begin{equation}
  \label{eq:Es_rx}
  (\boldsymbol{\mu}_{i,t}^{rx}, \boldsymbol{\sigma}_{i,t}^{rx}) = \mathcal{E}_s(\mathbf{w}_t, \boldsymbol{\mu}_{0,t}^{rx}, \boldsymbol{\sigma}_{0,t}^{rx}, \bar{\mathbf{y}}_{<i,t}^{rx}; \boldsymbol{\theta}_{\mathcal{E}_s}).
\end{equation}
The corresponding dequantized representation is obtained as:
\begin{equation}
  \label{eq:dequant_rx}
  \tilde{\mathbf{y}}_{i,t}^{rx}
  = \mathbf{w}_{i,t} \odot
  \left(\bar{\mathbf{y}}_{i,t}^{rx}+\boldsymbol{\mu}_{i,t}^{rx}\right),
  \quad \forall i\in\mathcal{I}.
\end{equation}

After merging $\{\tilde{\mathbf{y}}_{i,t}^{rx}\}_{i=0}^{N-1}$ into
$\tilde{\mathbf{y}}_t^{rx}$, the generator $\mathcal{G}(\cdot)$ 
reconstructs the video frame $\hat{\mathbf{x}}_t$ and the updated intermediate feature $\mathbf{F}_t^{rx}$ for the next time step:
\begin{equation}
  \label{eq:generate_rx}
  (\hat{\mathbf{x}}_t, \mathbf{F}_t^{rx})
  = \mathcal{G}\big(\tilde{\mathbf{y}}_t^{rx}
  \mid \mathbf{C}_{f,t-1}^{rx}; \boldsymbol{\theta}_{\mathcal{G}}\big).
\end{equation}

\subsubsection{Transmitter Buffer Update and Reference Alignment}
To maintain reference alignment for subsequent frames, PVSC updates the transmitter-side buffer using the receiver-side reconstruction path under an ideal-channel assumption, as shown in Fig.~\ref{Branch}(a). Specifically, the feature decoder $\mathcal{F}_{rx}(\cdot)$ takes $\mathbf{s}_t$, $\mathbf{k}_t$, and $\mathbf{C}_{s,t-1}^{tx}$ to produce $\bar{\mathbf{y}}_t^{tx}$ and update decoded state: 
\begin{equation}
  \label{eq:frx2}
 (\bar{\mathbf{y}}_t^{tx}, \mathbf{C}_{s,t}^{tx}) = \mathcal{F}_{rx}(\mathbf{s}_t,  \mathbf{k}_t \mid \mathbf{C}_{s,t-1}^{tx}; \boldsymbol{\theta}_{\mathcal{F}_{rx}}).
\end{equation}
Similarly, the initial parameter estimates and the spatial prior module $\mathcal{E}_s(\cdot)$ are used to compensate $\bar{\mathbf{y}}_t^{tx}$:
\begin{equation}
  \label{eq:Es_tx}
  (\boldsymbol{\mu}_{i,t}^{tx}, \boldsymbol{\sigma}_{i,t}^{tx}) = \mathcal{E}_s(\mathbf{w}_t, \boldsymbol{\mu}_{0,t}, \boldsymbol{\sigma}_{0,t}, \bar{\mathbf{y}}_{<i,t}^{tx}; \boldsymbol{\theta}_{\mathcal{E}_s}).
\end{equation}
The result is then fed into the generator $\mathcal{G}(\cdot)$ to update the intermediate feature $\mathbf{F}_t^{tx}$ for the next time step:
\begin{equation}
  \label{eq:generate_tx}
 (\hat{\mathbf{x}}_t^{tx}, \mathbf{F}_t^{tx}) = \mathcal{G}\big(\mathbf{w}_t \odot (\bar{\mathbf{y}}_t^{tx} + \boldsymbol{\mu}_{0,t}+ \sum_{i=1}\boldsymbol{\mu}_{i,t}^{tx}) \mid \mathbf{C}_{f,t-1}^{tx}; \boldsymbol{\theta}_{\mathcal{G}}\big).
\end{equation}

\subsection{Optimization Objective}
PVSC aims to transmit video with minimum bandwidth subject to a constraint on perceptual quality degradation. To reflect this goal, the overall loss is written in a rate-distortion (R-D) form that explicitly couples perceptual quality, adversarial realism, and bandwidth usage. Following~\cite{taming-transformers}, 
the reconstruction loss $\mathcal{L}_{\text{rec}}(\cdot)$ is defined as the $\ell_1$ distance between the input frame and its reconstruction. The perceptual loss $\mathcal{L}_{\text{per}}(\cdot)$~\cite{lpips}, e.g., LPIPS, measures high-level visual discrepancy, while the adversarial loss $\mathcal{L}_{\text{adv}}(\cdot)$~\cite{gan} promotes realistic textures and fine details. The rate loss $\mathcal{L}_{\text{rate}}(\cdot)$ penalizes both the average symbol length and the side-information overhead:
\begin{equation}
  \label{eq:Lrate}
  \begin{aligned}
    \mathcal{L}_{\text{rate}}(t)
    =\;& \sum_{i=1}^{H_y}\sum_{j=1}^{W_y} k_t^{(i,j)}
      + \eta\, \mathbb{E}_{\mathbf{k}_t}\big[ b(\mathbf{k}_t) \big] \\
      & - \eta\, 
      \mathbb{E}_{\mathbf{x}_t \sim p_{\mathbf{x}_t}}
      \mathbb{E}_{\mathbf{z}_t \sim p_{\mathbf{z}_t}}
      \big[ \log_{2} p_{\hat{\mathbf{z}}_t}(\hat{\mathbf{z}}_t) \big],
  \end{aligned}
\end{equation}
where $b(\mathbf{k}_t)$ is the number of bits required to encode the PNG-compressed rate map $\mathbf{k}_t$, and $\eta$ controls the spectral efficiency.

In practice, an ideal decoding branch is introduced, as shown in Fig.~\ref{Branch}(b). This branch generates $\hat{\mathbf{x}}_t^{ideal}$ without rate-aware symbol mapping:
\begin{equation}
  \label{eq:generate_ideal}
 (\hat{\mathbf{x}}_t^{ideal}, \mathbf{F}_t^{ideal}) = \mathcal{G}\big(\mathbf{w}_t \odot (\bar{\mathbf{y}}_t + \sum_{i=0}\boldsymbol{\mu}_{i,t}) \mid \mathbf{C}_{f,t-1}^{tx}; \boldsymbol{\theta}_{\mathcal{G}}\big).
\end{equation}
The transmitter and receiver branches produce $\hat{\mathbf{x}}_t^{tx}$ and $\hat{\mathbf{x}}_t$, respectively. All three reconstructions are optimized jointly to improve state consistency between the transmitter and receiver. 
Let $\mathcal{R}_t \triangleq \{\hat{\mathbf{x}}_t^{ideal}, \hat{\mathbf{x}}_t^{tx}, \hat{\mathbf{x}}_t\}$. The overall objective is:
\begin{equation}
  \label{eq:loss1}
  \begin{aligned}
    \mathcal{L}
    = \frac{1}{T} \sum_{t=1}^{T}\Bigg(
      &\omega_t \cdot \sum_{\mathbf{x}_t^{\mathrm{rec}} \in \mathcal{R}_t}
      \Big[
        \lambda_{\text{rec}} \,
        \mathcal{L}_{\text{rec}}\!\big(\mathbf{x}_t, \mathbf{x}_t^{\mathrm{rec}} \mid \mathbf{y}_t\big) \\
        &\quad\;\;+ \lambda_{\text{per}} \, \mathcal{L}_{\text{per}}\!\big(\mathbf{x}_t, \mathbf{x}_t^{\mathrm{rec}} \mid \mathbf{y}_t\big) \\
        &\quad\;\;
        + \lambda_{\text{adv}} \,
        \mathcal{L}_{\text{adv}}\!\big(\mathbf{x}_t, \mathbf{x}_t^{\mathrm{rec}} \mid \mathbf{y}_t\big)
      \Big] + \mathcal{L}_{\text{rate}}(t)
    \Bigg),
  \end{aligned}
\end{equation}
where $\omega_t$, $\lambda_{\text{rec}}$, $\lambda_{\text{per}}$, and $\lambda_{\text{adv}}$ are non-negative coefficients.

\begin{figure*}[htbp]
  \centering
  \includegraphics[width=1.0\textwidth]{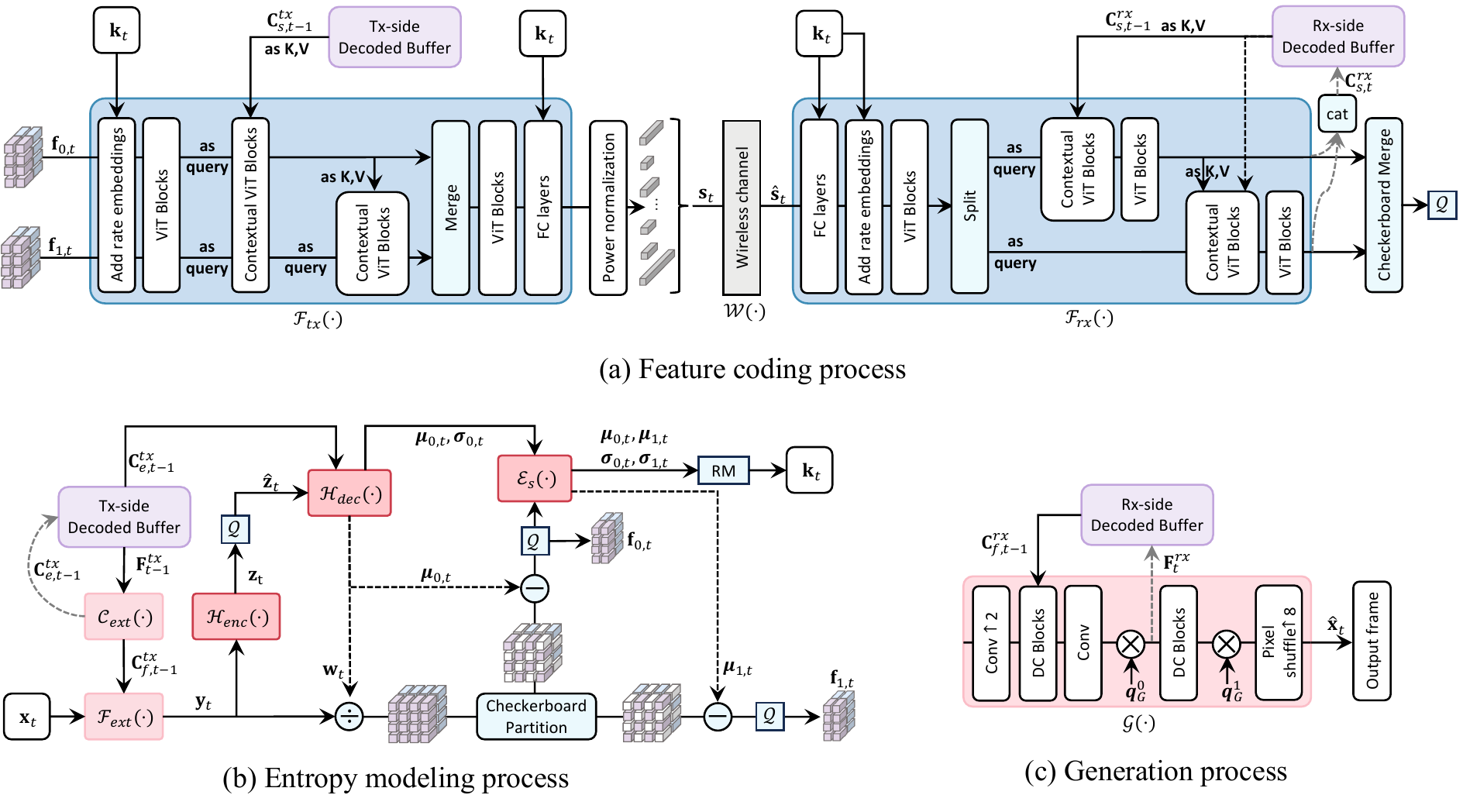}
  \caption{(a) Feature coding with ViT/contextual ViT blocks for spatial-temporal modeling and FC layers for complex-symbol projection. (b) Checkerboard-based spatio-temporal entropy modeling for predicting symbol-length factors from $\mathbf{y}_t$. (c) Generation of the reconstructed frame $\hat{\mathbf{x}}_t$ and updated intermediate feature $\mathbf{F}_t^{rx}$.}
  \label{process}
\end{figure*}

\section{Architecture and Implementations}
\label{sec:architecture}
This section details how the main methodological novelty of PVSC is implemented. We first instantiate the PVSC formulation in Section~\ref{sec:system_model} with concrete modules. Then, we introduce spatio-temporal feature coding with joint entropy modeling, followed by variable symbol-length mapping, rate-adaptive control, and the inference pipeline. Finally, we outline the GAN-based training strategy for end-to-end optimization.

\subsection{Network Architecture of PVSC}
Fig.~\ref{Branch}(b) shows the ideal decoding branch, which instantiates the context extractor $\mathcal{C}_{ext}(\cdot)$, feature extractor $\mathcal{F}_{ext}(\cdot)$, generator $\mathcal{G}(\cdot)$, spatial prior module $\mathcal{E}_s(\cdot)$, and hyperprior encoder–decoder pair $\mathcal{H}_{enc}(\cdot)$ and $\mathcal{H}_{dec}(\cdot)$. Before feature extraction, the input frame $\mathbf{x}_t$ is downsampled by pixel unshuffle, and processed by depthwise convolution blocks (DC blocks)~\cite{dcvc-dc, dcvc-fm, dcvc-rt} for compact representation. Subsequent feature extraction, initial parameter estimation, and frame generation are conditioned on the temporal context concatenated with the input features.

Fig.~\ref{process}(a) illustrates the feature coding pathway, which implements the feature encoder–decoder pair $\mathcal{F}_{tx}(\cdot)$ and $\mathcal{F}_{rx}(\cdot)$. 
We employ vision transformer (ViT) and contextual ViT blocks for feature coding, configured as in~\cite{Quaddeepsc}: the former model spatial dependencies in $\bar{\mathbf{y}}_t$ via self-attention, while the latter capture temporal dependencies through cross-attention to the previous decoded state and processed block. Position and rate embeddings are added to the ViT inputs, and fully connected (FC) layers map the features to symbols with the corresponding dimensions.

\subsection{Spatio-temporal Feature Coding}
PVSC performs joint design of entropy modeling and feature coding. The entropy model estimates the symbol length for each spatial unit $\bar{\mathbf{y}}_t^{(i,j)}$ and is aligned with the feature-coding path that maps $\bar{\mathbf{y}}_t^{(i,j)}$ to the channel symbol $\mathbf{s}_t^{(i,j)}$. The overall spatio-temporal entropy modeling process is illustrated in Fig.~\ref{process}(b). At the transmitter, the hyperprior decoder $\mathcal{H}_{dec}(\cdot)$ produces initial parameter estimates, including the quantization step $\mathbf{w}_t$, mean $\boldsymbol{\mu}_{0,t}$, and variance $\boldsymbol{\sigma}_{0,t}$. 
The semantic feature $\mathbf{y}_t$ is first normalized by $\mathbf{w}_t$ and then split into two checkerboard~\cite{he2021checkerboard} blocks $\{\mathbf{y}_{i,t}\}_{i=0}^1$ along the spatial dimensions. The first block $\mathbf{y}_{0,t}$ is centered with $\boldsymbol{\mu}_{0,t}$ and quantized as $\bar{\mathbf{y}}_{0,t}$, which is referred to as probabilistic feature quantization. Conditioned on $\bar{\mathbf{y}}_{0,t}$, the spatial prior module $\mathcal{E}_s(\cdot)$ infers the distribution parameters of the second block $\mathbf{y}_{1,t}$, \emph{i.e.}, $\boldsymbol{\mu}_{1,t}$ and $\boldsymbol{\sigma}_{1,t}$. 
The $\mathbf{y}_{1,t}$ is then centered with $\boldsymbol{\mu}_{1,t}$ and quantized as $\bar{\mathbf{y}}_{1,t}$.
Given $(\bar{\mathbf{y}}_{0,t}, \boldsymbol{\sigma}_{0,t})$ and $(\bar{\mathbf{y}}_{1,t}, \boldsymbol{\sigma}_{1,t})$, the spatial-unit symbol-length factor $k_t^{(i,j)}$ is computed according to~\eqref{eq:kij}, and its quantized indices constitute the rate map $\mathbf{k}_t$, thereby translating entropy-based redundancy estimation into fine-grained channel-symbol allocation.
Probabilistic feature dequantization is performed for both receiver-side reconstruction and transmitter-side buffer update.

The corresponding feature-coding path is shown in Fig.~\ref{process}(a). 
Before feature encoding, the two sparse checkerboard blocks $\bar{\mathbf{y}}_{0,t}$ and $\bar{\mathbf{y}}_{1,t}$ are compacted into dense feature blocks $\mathbf{f}_{0,t}$ and $\mathbf{f}_{1,t}$. 
This compaction removes empty spatial locations and reduces redundant computation. 
Inside $\mathcal{F}_{tx}(\cdot)$, the rate map $\mathbf{k}_t$ is used to select rate embeddings and the corresponding FC projection for each spatial unit. 
The blocks $\mathbf{f}_{0,t}$ and $\mathbf{f}_{1,t}$ are first processed by ViT blocks to model intra-block spatial dependencies and generate queries for contextual attention. 
The previous-step decoded state $\mathbf{C}_{s,t-1}^{tx}$ is used as keys and values in the contextual ViT blocks, thereby injecting temporal reference information without transmitting explicit motion vectors.
Cross-attention between the two block streams further captures inter-block spatial dependencies induced by the checkerboard partition. The attended blocks are merged and refined by a final ViT block to strengthen global spatial interactions.
The selected FC layers then project each spatial unit to complex channel symbols, producing $\mathbf{s}_t$ with a length matched to its estimated semantic entropy.
Consequently, PVSC establishes an entropy-aware and reference-conditioned feature-to-symbol mapping that suppresses spatio-temporal redundancy and improves robustness under bandwidth-limited wireless channels. 
At the receiver, the feature decoder $\mathcal{F}_{rx}(\cdot)$ applies the inverse rate-aware decoding procedure to recover $\bar{\mathbf{y}}_t^{rx}$ and updates the state $\mathbf{C}_{s,t}^{rx}$ for the next time step.

\subsection{Rate-Adaptive Mapping and Multi-Rate Control}
To improve the efficiency of channel-symbol allocation, PVSC constrains the symbol-length factor $k_t^{(i,j)}$ to a discrete rate set $\{k_i\}_{i=0}^{K-1}$. 
Each estimated $k_t^{(i,j)}$ is quantized to the nearest element in this set, and only the corresponding index is transmitted. 
To map each spatial unit $\bar{\mathbf{y}}_t^{(i,j)}$ from $C_y$ to $\mathcal{Q}(k_t^{(i,j)})=k_i$ dimensions, we apply a shared FC layer $\boldsymbol{FC}_i(\cdot)$ with weight shape $C_y \times C_y$, followed by a binary mask $\mathbf{m}_i$ for dimensionality reduction:
\begin{equation}
  \label{eq:eq18}
    \mathbf{r}_t^{(i,j)} = \boldsymbol{FC}_i(\mathbf{y}_{sp}^{(i,j)}; \boldsymbol{\theta}_{\boldsymbol{FC}_i}) \odot \mathbf{m}_i,
\end{equation}
where
\begin{equation}
  \label{eq:eq18-2}
  \mathbf{m}_i = [\underbrace{1,...,1}_{k_i}, 0,...,0 ] \in \mathbb{R}^{C_y}.
\end{equation}
The elements of $\mathbf{r}_t^{(i,j)}$ are then paired to form the complex symbol $\mathbf{s}_t^{(i,j)}$. The index of $\mathcal{Q}(k_t^{(i,j)})$ is also used to select the rate embedding, thereby coupling symbol-length assignment with rate-aware feature representation. 
At the beginning of training, the rate set is unrestricted and is initialized as $\{i\}_{i=0}^{C_y}$. 
After convergence, rarely activated rates are pruned, and only $K$ representative rate levels are retained. 
This pruning reduces redundant rate embeddings and FC layers, and reduces the search complexity of $k_t^{(i,j)}$, 
and improves the compactness of the rate map $\mathbf{k}_t$. As a result, $\mathbf{k}_t$ can be more efficiently compressed by lossless coding such as PNG, reducing the side-information overhead.

To support multi-rate transmission with a single trained model, PVSC introduces learnable channel-wise scaling vectors $\mathbf{q}_{ext}^{C}$, $\mathbf{q}_{ext}^{f}$, and $\{\mathbf{q}_{G}^{i}\}_{i=0}^{1}$, where all vectors lie in $\mathbb{R}^{C_y}$, as illustrated in Fig.~\ref{Branch}(b) and Fig.~\ref{process}(c). These vectors reweight the features in the context extractor $\mathcal{C}_{ext}(\cdot)$, the feature extractor $\mathcal{F}_{ext}(\cdot)$, and the generator $\mathcal{G}(\cdot)$, which provides lightweight rate adaptation and improves the flexibility of PVSC.

\begin{figure}[!t]
  \centering
  \includegraphics[width=3.4in]{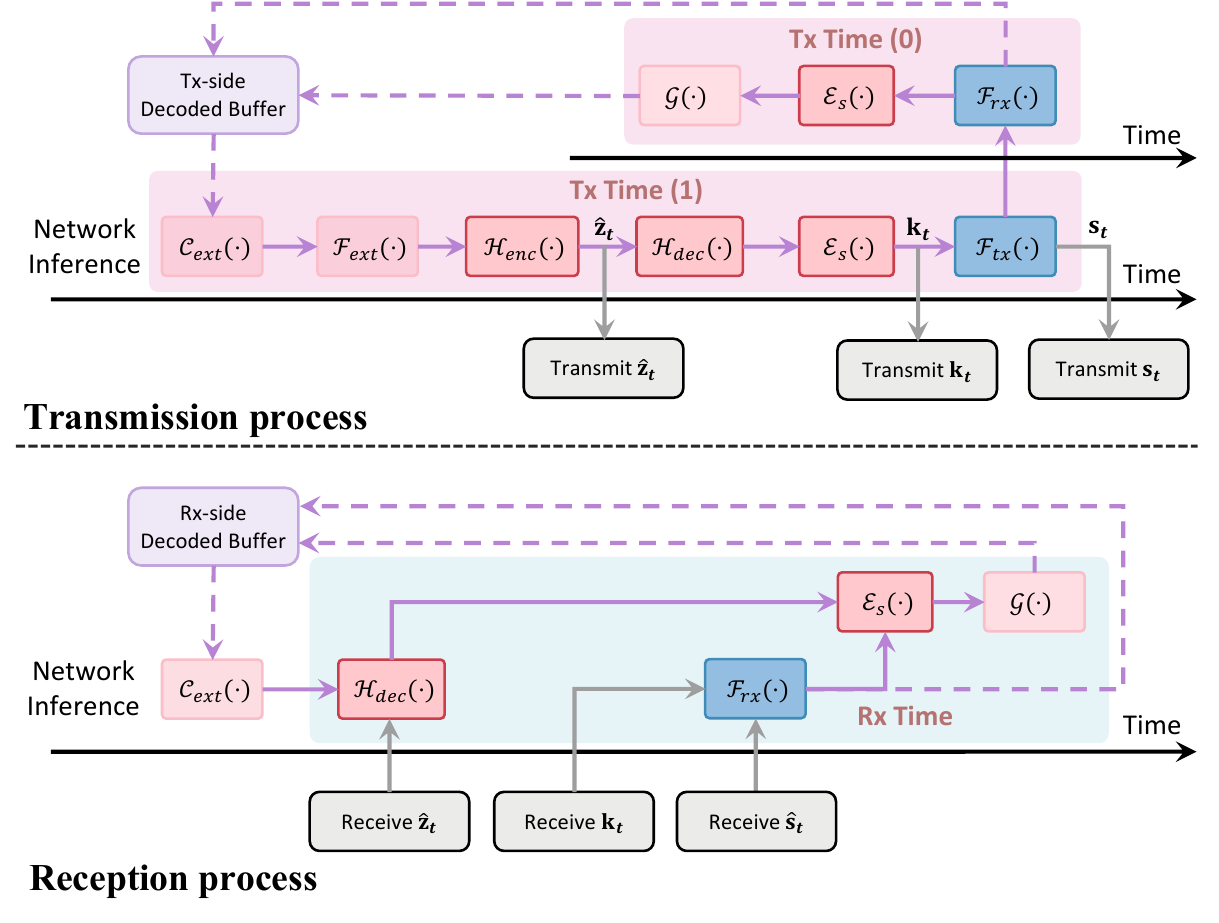}
  \caption{Transmission and reception pipeline.}
  \label{inference}
\end{figure}

Fig.~\ref{inference} summarizes the inference pipeline of PVSC. The rate-map bitstream $\mathbf{b}_t^{\mathbf{k}}$, the hyperprior bitstream $\mathbf{b}_t^{\mathbf{z}}$, and the channel symbols $\mathbf{s}_t$ are processed in a unified order at the transmitter and receiver. This design standardizes side-information handling and maintains temporal consistency between the transmitter and receiver reference buffers, enabling stable reconstruction across consecutive video frames.

\subsection{Training Methodology}
PVSC is trained using a three-stage procedure to stabilize the optimization of multiple coupled modules.
Jointly optimizing the semantic feature extraction, entropy modeling, rate-aware symbol mapping, and frame reconstruction modules from scratch causes unstable convergence. 
Therefore, training proceeds progressively from ideal reconstruction to channel-aware feature coding, and finally to end-to-end optimization.

In the first stage, the feature-coding modules $\mathcal{F}_{tx}(\cdot)$ and $\mathcal{F}_{rx}(\cdot)$ are frozen, and only the ideal decoding branch in Fig.~\ref{Branch}(b) is trained. This stage learns the semantic representation, entropy model, and generator without disturbance of wireless channel impairments. To stabilize early training, quantization is initially approximated using additive uniform noise~\cite{balle2016end}. After the reconstruction branch initially converges, quantization is implemented with the straight-through estimator (STE)~\cite{hu2023robust}, and GAN-related losses are introduced to reduce the train-test mismatch and improve perceptual quality.

In the second stage, the ideal decoding branch is fixed, while $\mathcal{F}_{tx}(\cdot)$ and $\mathcal{F}_{rx}(\cdot)$ are optimized by incorporating an explicit wireless channel into the training loop. This stage enables the feature encoder and decoder to learn channel-robust symbol mappings guided by the previously learned semantic and entropy models. An overcomplete rate set $\{i\}_{i=0}^{C_y}$ is retained in this stage, allowing the model to explore a wide range of symbol lengths per spatial unit.

In the third stage, all modules are unfrozen and jointly fine-tuned using a lower learning rate. After convergence, rarely activated rate levels are pruned from the initial set, retaining only the representative rate set ${k_i}{i=0}^{K-1}$. This rate-set distillation removes redundant rate embeddings and FC projections, reduces the search complexity of $k_t^{(i,j)}$, and improves the compressibility of the rate map $\mathbf{k}_t$. 

Across all stages, PVSC adopts a GOP-length curriculum: training starts with a short GOP for stable initial convergence and then gradually increases the GOP length to the target configuration. This curriculum improves the stability of temporal modeling and facilitates consistent reference-buffer learning across consecutive frames.

\section{Experimental Results}
\label{sec:experiments}
This section presents the experimental evaluation of PVSC. We first describe the experimental setup and then report results using both quantitative metrics and visual comparisons.

\begin{figure*}[htbp]
  \centering
  \includegraphics[width=1.0\textwidth]{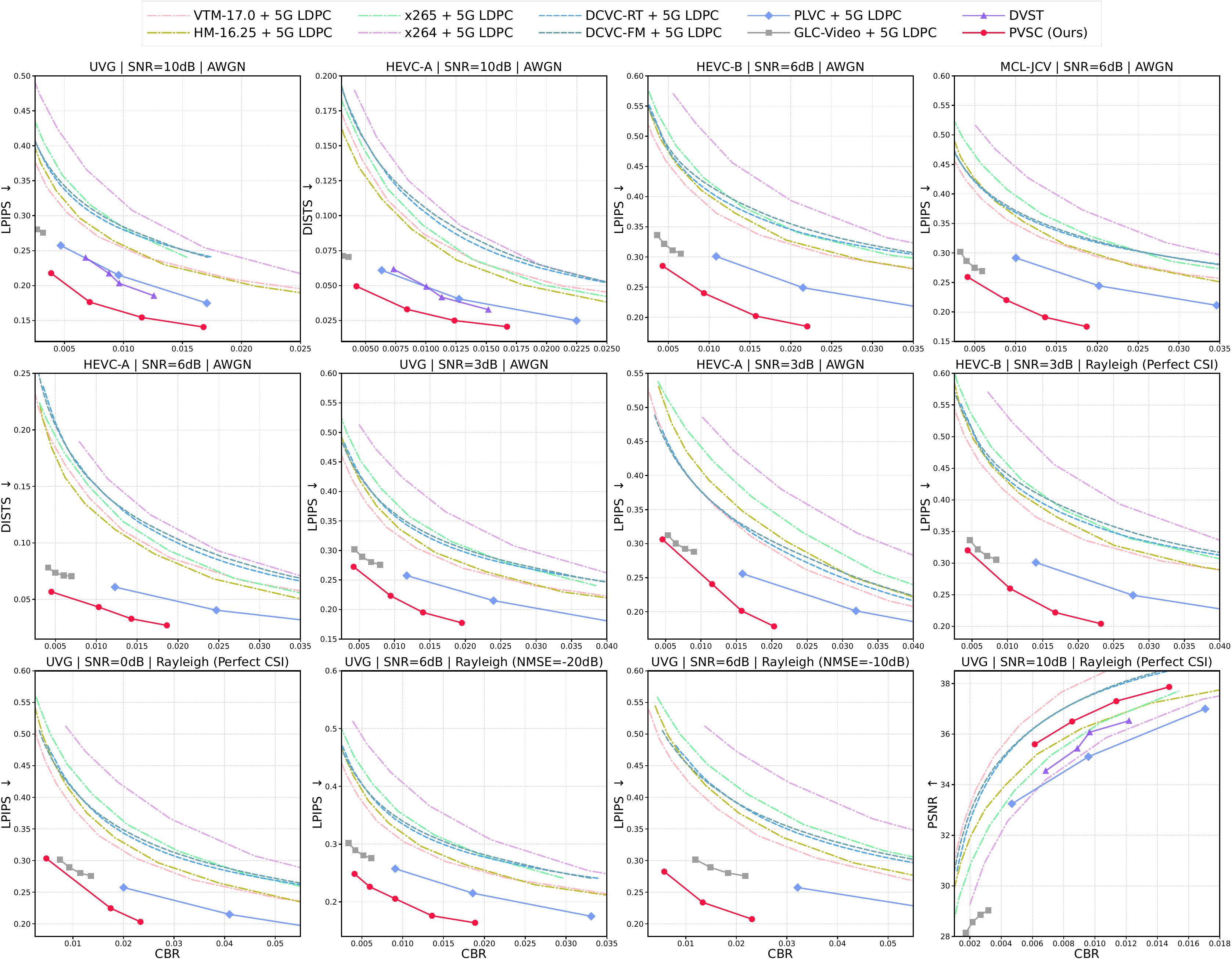}
  \caption{Rate-perception-distortion performance on 1080p/2K video datasets over different channels. The channel bandwidth ratio (CBR) denotes transmission cost, while lower LPIPS/DISTS and higher PSNR indicate better quality.}
  \label{r_d_p_result}
\end{figure*}

\subsection{Experimental Settings}
\paragraph{Training Details} 
PVSC is trained on Vimeo-90K~\cite{vimeo} and BVI-DVC~\cite{bvi-dvc}. 
All training frames are randomly cropped to $256 \times 256$. The GOP length is gradually increased to 7. 
PVSC is trained with multiple quality presets. 
The corresponding weights are set as $\omega_t \in {0.001, 0.10, 0.21, 0.42, 0.88}$ for different learnable channel-wise scaling vectors. 
The lagrange multipliers $\lambda_{\text{rec}}$, $\lambda_{\text{per}}$, and $\lambda_{\text{adv}}$ are set to $1.0$, $0.8$, and $0.1$, respectively. 
The scaling hyperparameter $\eta$ is set to $0.2$. 
The channel dimensions are fixed as $C_y = 128$ and $C_z = 128$. 
Adam optimizer is used with $\beta_1 = 0.9$ and $\beta_2 = 0.999$,
an initial learning rate of $1 \times 10^{-4}$, and a batch size of 8. 
During end-to-end fine-tuning, the learning rate is reduced to $1 \times 10^{-5}$. 
The initial rate set $\{i\}_{i=0}^{C_y}$ is pruned to $\{0, 4, 8, 12, 16, 20, 24, 28, 36, 44, 52, 60, 68, 84, 100, 128\}$. 
The model is first trained over an AWGN channel at SNR $=10$~dB; models for other channel conditions are initialized from this checkpoint and then fine-tuned. 
Training is conducted on an NVIDIA A100 GPU, while evaluation is performed on an NVIDIA GeForce RTX 4090 GPU.

\begin{table*}[t]
  \centering
  \caption{BD-CBR (\%) on LPIPS / DISTS with respect to VTM-17.0 + 5G LDPC over different datasets and GOPs, evaluated over an AWGN channel with SNR = 6 dB. Smaller values indicate larger bandwidth savings at comparable perceptual quality. The best and second-best results are shown in $\textbf{bold}$ and $\underline{\text{underlined}}$, respectively.}
  \renewcommand{\arraystretch}{1.2}
  \setlength{\tabcolsep}{3.5pt}
  \footnotesize
  \begin{tabular}{l|c|ccccccc|c}
    \toprule
    \textbf{Method} & \textbf{GOP} &
    \textbf{UVG} & \textbf{MCL-JCV} & \textbf{HEVC A} &
    \textbf{HEVC B} & \textbf{HEVC C} & \textbf{HEVC D} & \textbf{HEVC E} &
    \textbf{Average} \\
    \midrule
    \multirow{1}{*}{VTM-17.0 + 5G LDPC}
      & -  & 0.0 / 0.0 & 0.0 / 0.0 & 0.0 / 0.0 & 0.0 / 0.0 & 0.0 / 0.0 & 0.0 / 0.0 & 0.0 / 0.0 & 0.0 / 0.0 \\
    \midrule
    \multirow{3}{*}{HM-16.25 + 5G LDPC}
      & 4  & 12.1 / -17.6 & 13.0 / -19.1 & 21.2 / -10.6 & 12.4 / -17.3 & 15.5 / -4.1 & 16.8 / 5.1 & 16.1 / -17.1 & 15.3 / -11.5 \\
      & 8  & 10.3 / -23.6 & 14.1 / -22.7 & 22.0 / -13.8 & 10.9 / -23.1 & 16.2 / -5.7 & 18.7 / 6.9 & 14.6 / -19.8 & 15.3 / -14.5 \\
      & 12 & 10.0 / -26.9 & 14.4 / -24.9 & 22.8 / -15.1 & 10.3 / -25.7 & 16.7 / -7.7 & 20.0 / 6.6 & 13.8 / -22.3 & 15.4 / -16.6 \\
    \midrule
    \multirow{3}{*}{x265 (veryslow) + 5G LDPC}
      & 4  & 70.5 / 16.8 & 79.4 / 11.5 & 53.9 / 8.9 & 54.1 / 2.9 & 44.3 / 3.7 & 31.8 / 11.7 & 51.1 / -6.5 & 55.0 / 7.0 \\
      & 8  & 96.5 / 23.1 & 111.2 / 22.2 & 70.9 / 14.8 & 68.0 / 7.6 & 59.7 / 9.2 & 46.9 / 24.1 & 60.8 / -2.0 & 73.4 / 14.1 \\
      & 12 & 107.0 / 20.9 & 118.8 / 20.2 & 72.6 / 13.5 & 75.0 / 7.5 & 66.6 / 8.0 & 59.9 / 31.4 & 64.1 / -2.2 & 80.6 / 14.2 \\
    \midrule
    \multirow{3}{*}{x264 (veryslow) + 5G LDPC}
      & 4  & 95.6 / 29.8 & 92.0 / 31.4 & 77.1 / 35.1 & 68.8 / 20.5 & 70.6 / 47.0 & 64.5 / 56.7 & 123.9 / 50.1 & 84.7 / 38.6 \\
      & 8  & 119.5 / 39.8 & 120.0 / 45.1 & 90.1 / 41.9 & 83.1 / 28.9 & 86.0 / 55.9 & 82.4 / 72.8 & 138.5 / 58.5 & 102.8 / 49.0 \\
      & 12 & 126.2 / 38.0 & 126.5 / 47.0 & 90.1 / 41.2 & 91.0 / 31.6 & 95.2 / 63.7 & 98.6 / 83.0 & 139.9 / 57.5 & 109.7 / 51.7 \\
    \midrule
    \multirow{3}{*}{DCVC-RT + 5G LDPC}
      & 4  & 40.8 / 67.3 & 31.3 / 59.0 & 5.9 / 31.0 & 36.1 / 54.1 & 16.6 / 41.3 & 2.2 / 15.1 & 28.4 / 43.4 & 23.0 / 44.4 \\
      & 8  & 42.6 / 72.5 & 35.5 / 67.3 & -1.5 / 22.2 & 37.9 / 61.5 & 11.2 / 37.3 & 1.8 / 17.6 & 34.9 / 50.3 & 23.2 / 47.0 \\
      & 12 & 37.6 / 65.5 & 28.8 / 60.9 & -10.0 / 11.7 & 35.2 / 58.1 & 5.3 / 28.0 & -0.6 / 12.4 & 32.8 / 45.4 & 18.4 / 40.3 \\
    \midrule
    \multirow{3}{*}{DCVC-FM + 5G LDPC}
      & 4  & 41.2 / 66.4 & 32.4 / 61.2 & 7.1 / 33.7 & 43.3 / 62.8 & 14.4 / 42.9 & 3.4 / 18.2 & 30.2 / 51.6 & 24.6 / 48.1 \\
      & 8  & 41.0 / 71.8 & 37.6 / 70.9 & -1.5 / 25.9 & 41.6 / 66.0 & 6.9 / 37.8 & -0.2 / 17.3 & 28.1 / 50.8 & 21.9 / 48.6 \\
      & 12 & 43.0 / 74.8 & 35.6 / 69.5 & -6.4 / 19.9 & 43.8 / 69.1 & 3.4 / 30.5 & -1.2 / 13.2 & 31.5 / 53.2 & 21.4 / 47.2 \\
    \midrule
    \multirow{3}{*}{PLVC + 5G LDPC}
      & 4  & -46.8 / -85.9 & -51.3 / -79.2 & -25.8 / -56.3 & -61.7 / -86.1 & -5.5 / -27.2 & 35.9 / 22.4 & \underline{-75.4} / -73.9 & -33.0 / -55.2 \\
      & 8  & -19.1 / -80.5 & -28.9 / -72.0 & -5.2 / -35.2 & -38.1 / -81.6 & 17.2 / -7.4 & 58.8 / 41.2 & -59.6 / -54.6 & -10.7 / -41.5 \\
      & 12 & 35.6 / -60.6 & 10.9 / -53.6 & 24.5 / -7.3 & 2.3 / -68.0 & 46.0 / 16.4 & 90.7 / 62.8 & -7.6 / -13.9 & 28.9 / -17.7 \\
    \midrule
    \multirow{3}{*}{GLC-Video + 5G LDPC}
      & 4  & \underline{-64.2} / \underline{-91.4} & \underline{-81.5} / \textbf{-93.4} & \underline{-64.4} / \underline{-79.4} & \underline{-76.3} / \underline{-94.1} & \underline{-69.1} / \underline{-79.5} & \underline{-64.4} / \underline{-77.6} & -73.9 / \underline{-79.0} & \underline{-70.5} / \underline{-84.9} \\
      & 8  & \underline{-61.0} / \underline{-92.3} & \underline{-79.7} / \textbf{-93.4} & \underline{-62.4} / \underline{-79.6} & \underline{-72.9} / \underline{-93.6} & \underline{-66.2} / \underline{-76.5} & \underline{-59.8} / \underline{-74.9} & \underline{-70.9} / \underline{-77.0} & \underline{-67.6} / \underline{-83.9} \\
      & 12 & \underline{-56.8} / \underline{-92.2} & \textbf{-78.2} / \textbf{-93.1} & \underline{-60.2} / \underline{-78.3} & \underline{-69.1} / \underline{-93.4} & \underline{-62.3} / \underline{-75.2} & \underline{-53.6} / \underline{-71.4} & \underline{-64.6} / \underline{-72.3} & \underline{-63.6} / \underline{-82.3} \\
    \midrule
    \multirow{3}{*}{PVSC (Ours)}
      & 4  & \textbf{-82.7} / \textbf{-94.3} & \textbf{-87.0} / \underline{-93.0} & \textbf{-73.6} / \textbf{-85.7} & \textbf{-87.6} / \textbf{-94.3} & \textbf{-76.6} / \textbf{-84.8} & \textbf{-73.6} / \textbf{-78.2} & \textbf{-92.9} / \textbf{-92.1} & \textbf{-82.0} / \textbf{-88.9} \\
      & 8  & \textbf{-81.4} / \textbf{-96.0} & \textbf{-82.7} / \underline{-91.3} & \textbf{-72.2} / \textbf{-85.8} & \textbf{-85.4} / \textbf{-94.8} & \textbf{-73.1} / \textbf{-83.3} & \textbf{-68.8} / \textbf{-75.6} & \textbf{-94.9} / \textbf{-94.7} & \textbf{-79.8} / \textbf{-88.8} \\
      & 12 & \textbf{-80.1} / \textbf{-95.8} & \underline{-77.8} / \underline{-90.2} & \textbf{-74.7} / \textbf{-84.3} & \textbf{-80.1} / \textbf{-94.4} & \textbf{-71.1} / \textbf{-80.7} & \textbf{-62.2} / \textbf{-75.4} & \textbf{-94.9} / \textbf{-94.8} & \textbf{-77.3} / \textbf{-87.9} \\
    \bottomrule
  \end{tabular}
  \label{tab:bd_rate_lpips_dists}
\end{table*}

\paragraph{Evaluation Settings} 
PVSC is evaluated using LPIPS and DISTS, where LPIPS is computed with VGG features~\cite{vgg}. 
Bandwidth efficiency is measured using the Bj{\o}ntegaard-Delta CBR (BD-CBR) metric~\cite{bjontegaard2001calculation}, 
where negative values indicate CBR savings. 
The test datasets include HEVC classes A–E~\cite{hevcclass}, UVG~\cite{uvg}, and MCL-JCV~\cite{mcl-jcv}, covering resolutions from 240p to 2K and representing typical wireless video transmission scenarios. 
All frames are padded to multiples of 128 pixels for evaluation. The raw test sequences are stored in YUV420 format. 
Following the test conditions in~\cite{dcvc-rt, glc-video}, we convert YUV420 to RGB using BT.709, and further convert RGB frames to 10-bit YUV444 as inputs to traditional video codecs. 

The traditional video codecs include x264/FFmpeg~\cite{FFmpeg}, x265/FFmpeg~\cite{FFmpeg}, H.265/HM~\cite{hevc_hm}, and H.266/VTM~\cite{vvc_vtm}. For x264 and x265, we use the zero-latency mode and the veryslow preset. For HM (version 16.25) and VTM (version 17.0), we adopt lowdelay-P configurations. For each sequence, three consecutive GOPs are encoded to compute the average reconstruction quality and CBR. 
Abnormally high-rate operating points are treated as outliers and excluded from BD-CBR computation.
This setup approximates low-latency transmission conditions for traditional video codecs. The LVCs include GLC-Video~\cite{glc-video}, PLVC~\cite{plvc}, DCVC-FM~\cite{dcvc-fm}, and DCVC-RT~\cite{dcvc-rt}. 
For PLVC, P-frames and bi-directionally coded frames each account for half of the frames within each GOP.
The video semantic communication baseline is DVST~\cite{dvst}. We follow the official implementations of all the above learned baselines and evaluate them on RGB frames. 
For a fair comparison, DVST and PVSC are trained with the GAN-based loss in~\eqref{eq:loss1} for LPIPS/DISTS-CBR evaluation, and with a distortion-oriented loss for PSNR-CBR evaluation.  

Each traditional codec and LVC baseline is combined with 5G LDPC channel coding~\cite{richardson2018design} and digital modulation following the 3GPP specification, with a block length of 6144. 
The modulation order and code rate are selected from MCS Table 5.1.3.1-1 for the PDSCH in~\cite{3gpp38214}. For each channel condition, we sweep the MCS index and report the best-performing configuration. 
All simulations are implemented in Sionna~\cite{sionna}. Source and channel coding schemes are concatenated with the ``+" symbol. For the side link, the bitstreams $\mathbf{b}_t^{\mathbf{k}}$ and $\mathbf{b}_t^{\mathbf{z}}$ are transmitted using the same configuration as the corresponding baseline. For I-frames, we use the model in~\cite{Quaddeepsc} and fine-tune it with the GAN-based loss in~\eqref{eq:loss1}. The P-frames are transmitted using the proposed PVSC framework.

\subsection{Results Analysis}
\paragraph{Transmission Efficiency and Robustness}
Fig.~\ref{r_d_p_result} presents the rate-perception-distortion performance of different schemes on high-resolution video datasets under various channel conditions. 
It can be observed that PVSC consistently achieves more favorable performance than existing open-source MSE-optimized LVCs, perceptual LVCs, and traditional video codecs protected by 5G LDPC channel coding. 
In particular, PVSC achieves lower LPIPS and DISTS values at substantially reduced CBRs, especially under short-blocklength transmission where separate source-channel coding systems suffer from redundancy overhead.

To quantify transmission efficiency, Table~\ref{tab:bd_rate_lpips_dists} reports the BD-CBR results under the AWGN channel at SNR = 6 dB, using ``VTM + 5G LDPC'' as the anchor. 
The results cover different coding schemes, datasets, and GOP configurations, and show that PVSC achieves substantial average CBR savings over seven datasets in terms of both LPIPS and DISTS, demonstrating its high transmission efficiency.

We further evaluate the robustness of PVSC under diverse channel conditions. 
Table~\ref{tab:bd_rate_lpips_dists_SNRs} reports the BD-CBR results at different AWGN SNR levels, while Table~\ref{tab:bd_rate_rayleigh_imperfect_csi_pvsc} extends the evaluation to a Rayleigh fading channel with $L_{co}=1$ and imperfect CSI, characterized by $\mathrm{NMSE}_{\mathrm{dB}}$. 
In the Rayleigh fading case, PVSC employs ZF equalization as in~\eqref{eq:zfrx}, whereas the competing baselines use minimum mean-square error (MMSE) equalization. 
Together with the performance curves in Fig.~\ref{r_d_p_result}, these results show that PVSC consistently maintains substantial gains across different channel conditions, demonstrating its robustness to varying transmission environments.

\begin{table*}[t]
  \centering
  \caption{BD-CBR (\%) on LPIPS / DISTS with respect to VTM-17.0 + 5G LDPC over different datasets and GOPs, evaluated over an AWGN channel with different SNRs. Smaller values indicate larger bandwidth savings at comparable perceptual quality.}
  \renewcommand{\arraystretch}{1.2}
  \setlength{\tabcolsep}{3.5pt}
  \footnotesize
  \begin{tabular}{l|c|ccccccc|c}
    \toprule
    \textbf{Method} & \textbf{GOP} &
    \textbf{UVG} & \textbf{MCL-JCV} & \textbf{HEVC A} &
    \textbf{HEVC B} & \textbf{HEVC C} & \textbf{HEVC D} & \textbf{HEVC E} &
    \textbf{Average} \\
    \midrule
    \multirow{1}{*}{VTM-17.0 + 5G LDPC}
      & --  & 0.0 / 0.0 & 0.0 / 0.0 & 0.0 / 0.0 & 0.0 / 0.0 & 0.0 / 0.0 & 0.0 / 0.0 & 0.0 / 0.0 & 0.0 / 0.0 \\
    \midrule
    \multirow{3}{*}{DVST (SNR = 10 dB)}
      & 4  & -50.3 / -79.5 & -60.6 / -75.6 & -34.6 / -55.7 & -65.8 / -81.0 & -44.5 / -57.8 & -49.3 / -55.3 & -67.8 / -65.8 & -53.3 / -67.2 \\
      & 8  & -38.8 / -78.9 & -48.4 / -73.3 & -24.2 / -50.8 & -50.3 / -76.9 & -34.3 / -53.6 & -41.1 / -51.4 & -60.9 / -63.8 & -42.6 / -64.1 \\
      & 12 & -0.8 / -58.8  & -23.6 / -58.4 & -10.1 / -34.8 & -9.6 / -46.9  & -7.3 / -31.5  & -22.3 / -40.0 & -10.4 / -22.3 & -12.0 / -41.8 \\
    \midrule
    \multirow{3}{*}{PVSC (SNR = 10 dB)}
      & 4  & -78.1 / -92.2 & -83.3 / -90.4 & -65.4 / -80.0 & -83.8 / -91.7 & -68.1 / -78.9 & -62.1 / -71.1 & -90.2 / -89.6 & -75.9 / -84.8 \\
      & 8  & -79.2 / -93.8 & -78.9 / -88.6 & -62.6 / -80.8 & -83.1 / -92.3 & -66.7 / -77.3 & -59.7 / -71.0 & -93.1 / -92.6 & -74.8 / -85.2 \\
      & 12 & -75.0 / -94.1 & -74.4 / -86.8 & -67.5 / -81.5 & -78.7 / -91.5 & -57.5 / -74.9 & -54.1 / -68.8 & -93.8 / -94.1 & -71.6 / -84.5 \\
    \midrule
    \multirow{3}{*}{PVSC (SNR = 6 dB)}
      & 4  & -82.7 / -94.3 & -87.0 / -93.0 & -73.6 / -85.7 & -87.6 / -94.3 & -76.6 / -84.8 & -73.6 / -78.2 & -92.9 / -92.1 & -82.0 / -88.9 \\
      & 8  & -81.4 / -96.0 & -82.7 / -91.3 & -72.2 / -85.8 & -85.4 / -94.8 & -73.1 / -83.3 & -68.8 / -75.6 & -94.9 / -94.7 & -79.8 / -88.8 \\
      & 12 & -80.1 / -95.8 & -77.8 / -90.2 & -74.7 / -84.3 & -80.1 / -94.4 & -71.1 / -80.7 & -62.2 / -75.4 & -94.9 / -94.8 & -77.3 / -87.9 \\
    \midrule
    \multirow{3}{*}{PVSC (SNR = 3 dB)}
      & 4  & -78.1 / -93.6 & -84.3 / -93.6 & -68.7 / -84.8 & -85.0 / -94.6 & -72.3 / -83.0 & -67.9 / -76.4 & -91.7 / -89.3 & -78.3 / -87.9 \\
      & 8  & -72.4 / -93.8 & -77.3 / -92.0 & -66.7 / -85.5 & -80.2 / -93.7 & -68.4 / -81.5 & -60.5 / -73.0 & -92.8 / -91.7 & -74.0 / -87.3 \\
      & 12 & -66.8 / -92.6 & -70.4 / -89.0 & -64.9 / -85.6 & -73.4 / -92.3 & -59.3 / -81.6 & -53.7 / -72.1 & -92.7 / -92.0 & -68.7 / -86.5 \\
    \midrule
    \multirow{3}{*}{PVSC (SNR = 0 dB)}
      & 4  & -76.6 / -92.9 & -84.2 / -93.7 & -72.7 / -86.2 & -83.3 / -94.3 & -75.3 / -84.7 & -68.0 / -78.3 & -90.0 / -87.0 & -78.6 / -88.1 \\
      & 8  & -74.8 / -93.1 & -77.0 / -91.5 & -74.1 / -87.1 & -78.4 / -93.6 & -68.0 / -82.5 & -58.8 / -74.0 & -90.7 / -87.6 & -74.5 / -87.1 \\
      & 12 & -69.0 / -92.3 & -70.7 / -89.3 & -69.6 / -86.6 & -71.2 / -91.7 & -61.0 / -81.6 & -49.6 / -71.3 & -89.4 / -87.7 & -68.6 / -85.8 \\
    \bottomrule
  \end{tabular}
  \label{tab:bd_rate_lpips_dists_SNRs}
\end{table*}

\begin{table*}[t]
  \centering
  \caption{BD-CBR (\%) on LPIPS / DISTS with respect to VTM-17.0 + 5G LDPC over different datasets, evaluated over a Rayleigh fading channel with imperfect CSI, SNR = 6 dB, and GOP = 4. Smaller values indicate larger bandwidth savings at comparable perceptual quality.}
  \renewcommand{\arraystretch}{1.2}
  \setlength{\tabcolsep}{3.2pt}
  \footnotesize
  \begin{tabular}{l|c|ccccccc|c}
    \toprule
    \textbf{Method} & \textbf{CSI NMSE} &
    \textbf{UVG} & \textbf{MCL-JCV} & \textbf{HEVC A} &
    \textbf{HEVC B} & \textbf{HEVC C} & \textbf{HEVC D} & \textbf{HEVC E} &
    \textbf{Average} \\
    \midrule
    \multirow{1}{*}{VTM-17.0 + 5G LDPC}
      & - & 0.0 / 0.0 & 0.0 / 0.0 & 0.0 / 0.0 & 0.0 / 0.0 & 0.0 / 0.0 & 0.0 / 0.0 & 0.0 / 0.0 & 0.0 / 0.0 \\
    \midrule
    \multirow{2}{*}{PVSC (Ours)}
      & -20 dB & -82.7 / -94.3 & -87.0 / -93.0 & -73.6 / -85.7 & -87.6 / -94.3 & -76.6 / -84.8 & -73.6 / -78.2 & -92.9 / -92.1 & -82.0 / -88.9 \\
      & -10 dB & -95.1 / -98.4 & -96.3 / -98.0 & -92.5 / -96.0 & -96.5 / -98.4 & -93.4 / -95.7 & -92.5 / -93.8 & -98.0 / -97.8 & -94.9 / -96.9 \\
    \bottomrule
  \end{tabular}
  \label{tab:bd_rate_rayleigh_imperfect_csi_pvsc}
\end{table*}

\begin{figure*}[htbp]
  \centering
  \includegraphics[width=1.0\textwidth]{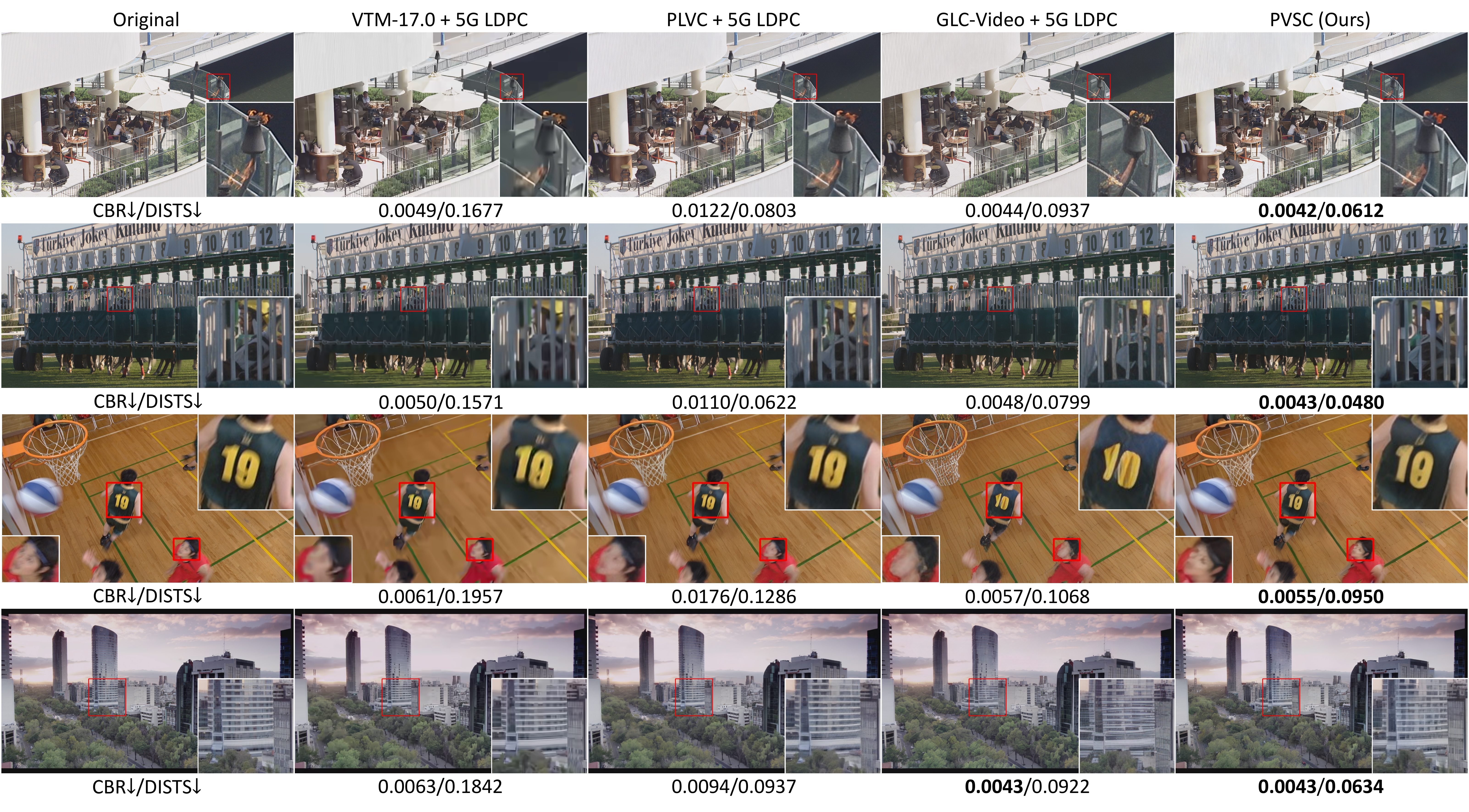}
  \caption{Visual comparison of different methods over an AWGN channel (SNR = 6 dB). Zoom in for a better view.}
  \label{Visualize}
\end{figure*}

\begin{table*}[t]
  \centering
  \caption{
    Computational complexity comparison at different video resolutions on an NVIDIA RTX~4090 GPU.
    For PVSC and DVST, we report the end-to-end transmitter and receiver latency (Tx/Rx time).
    For PVSC, the Tx/Rx time is measured by following the inference pipeline in Fig.~\ref{inference}. 
    For GLC-Video, only source-coding latency is reported, excluding channel coding. 
    MACs and parameter counts are also reported, where the parameter count corresponds to the model parameters required to support 4 output rates.
    The best and second-best results are shown in \textbf{bold} and \underline{underline}, respectively.}
  \label{tab:merged_complexity}
  \renewcommand{\arraystretch}{0.95}
  \setlength{\tabcolsep}{4.0pt}
  \footnotesize
  \resizebox{\textwidth}{!}{
  \begin{tabular}{@{}c cccc cccc cccc@{}}
    \toprule
    \multirow{2}{*}{\textbf{Resolution}} &
    \multicolumn{4}{c}{\textbf{PVSC (Ours)}} &
    \multicolumn{4}{c}{\textbf{DVST}} &
    \multicolumn{4}{c}{\textbf{GLC-Video}} \\
    \cmidrule(lr){2-5}
    \cmidrule(lr){6-9}
    \cmidrule(lr){10-13}
    &
    \makecell{\textbf{Tx time}\\\textbf{(ms)}} &
    \makecell{\textbf{Rx time}\\\textbf{(ms)}} &
    \makecell{\textbf{MACs}\\\textbf{(M/pixel)}} &
    \makecell{\textbf{Params}\\\textbf{(M)}} &
    \makecell{\textbf{Tx time}\\\textbf{(ms)}} &
    \makecell{\textbf{Rx time}\\\textbf{(ms)}} &
    \makecell{\textbf{MACs}\\\textbf{(M/pixel)}} &
    \makecell{\textbf{Params}\\\textbf{(M)}} &
    \makecell{\textbf{Src. Enc.}\\\textbf{(ms)}} &
    \makecell{\textbf{Src. Dec.}\\\textbf{(ms)}} &
    \makecell{\textbf{MACs}\\\textbf{(M/pixel)}} &
    \makecell{\textbf{Params}\\\textbf{(M)}} \\
    \midrule
    2K
    & \textbf{74.0} & \textbf{44.7} & \textbf{0.41} & \textbf{35.4}
    & 373.9 & \underline{128.9} & \underline{2.60} & \underline{12.1$\times$4}
    & \underline{191.8} & 303.2 & 3.26 & 125.3 \\

    1080p
    & \textbf{44.1} & \textbf{27.0} & \textbf{0.41} & \textbf{35.4}
    & 351.8 & \underline{122.1} & \underline{2.77} & \underline{12.1$\times$4}
    & \underline{167.8} & 266.9 & 3.28 & 125.3 \\

    720p
    & \textbf{22.2} & \textbf{13.5} & \textbf{0.41} & \textbf{35.4}
    & 157.9 & \underline{54.7} & \underline{2.77} & \underline{12.1$\times$4}
    & \underline{79.3} & 125.0 & 3.48 & 125.3 \\

    480p
    & \textbf{19.3} & \textbf{11.0} & \textbf{0.43} & \textbf{35.4}
    & 74.1 & \underline{25.4} & \underline{2.91} & \underline{12.1$\times$4}
    & \underline{37.3} & 58.0 & 3.65 & 125.3 \\
    \bottomrule
  \end{tabular}
  }
\end{table*}

\paragraph{Computational Complexity}
In practical deployments, computational complexity and inference latency are critical factors for real-time video transmission on target hardware platforms. 
Table~\ref{tab:merged_complexity} compares PVSC with DVST and GLC-Video at different video resolutions on an NVIDIA RTX~4090 GPU. 
For PVSC and DVST, we report the end-to-end transmitter and receiver latency, whereas GLC-Video is evaluated only as a source codec without channel-coding latency. 
Despite this favorable setting for GLC-Video, PVSC achieves the lowest latency and the smallest MACs per pixel across all resolutions.

Specifically, PVSC reaches 22.7/37.0~FPS, 45.0/74.1~FPS, and 51.8/90.9~FPS at the transmitter/receiver for 1080p, 720p, and 480p videos, respectively. 
Meanwhile, its computational cost remains only 0.41--0.43~M MACs/pixel, which is much lower than those of DVST and GLC-Video. 
These results show that, although perceptual LVCs such as GLC-Video can improve visual quality, they often incur substantially higher complexity and latency. 
In contrast, PVSC provides a more favorable tradeoff between perceptual transmission performance and computational efficiency, making it suitable for various-resolution, low-latency video streaming over bandwidth-limited wireless links.

\paragraph{Visual Perceptual Quality}
Fig.~\ref{Visualize} compares reconstructed frames over an AWGN channel at SNR = 6 dB in the low-CBR regime (short-blocklength), with regions of interest zoomed in for detailed inspection. 
PVSC achieves the lowest CBR while preserving sharper textures and structures than the competing baselines, which often suffer from blurring and structural distortions in fine-grained regions. 
The visual improvements are consistent with the lower DISTS values, demonstrating the effectiveness of PVSC in delivering perceptually pleasing high-resolution video under stringent bandwidth constraints.

\subsection{Ablation Studies}

We conduct ablation studies to assess the proposed spatio-temporal feature coding module and loss design. 
Table~\ref{tab:ablation_bd} reports the BD-CBR and complexity of different variants, where $M_e$ denotes the full PVSC system. 
Removing spatio-temporal feature coding yields $M_a$, which reduces MACs and latency but substantially increases BD-CBR. 
The temporal-only variant $M_b$ and spatial-only variant $M_c$ both improve over $M_a$ but remain inferior to $M_e$, confirming the complementarity of temporal and spatial feature coding. 
The resulting increase in MACs and latency also reflects the tradeoff between bandwidth saving and computational complexity. 
Comparing $M_d$ with $M_e$ further shows that rate-set distillation reduces BD-CBR, model size, and inference latency without additional computational cost, thereby improving deployment efficiency.

Table~\ref{tab:loss_ablation_psnr_lpips_dists} evaluates the contribution of different loss terms. 
Using only $\mathcal{L}_{\mathrm{rec}}$ favors PSNR-oriented fidelity but leads to inferior perceptual efficiency in terms of LPIPS and DISTS. 
Adding $\mathcal{L}_{\mathrm{per}}$ significantly improves perceptual BD-CBR, while $\mathcal{L}_{\mathrm{adv}}$ brings further gains. 
With the full loss design, PVSC achieves the best LPIPS and DISTS BD-CBR values of $-78.1\%$ and $-92.2\%$, respectively. 
Although DVST adopts the same loss configuration, PVSC consistently achieves better performance, further validating the proposed architecture.

\begin{table}[t]
  \centering
  \caption{Ablation study on the UVG dataset (1080p) over an AWGN channel (SNR = 10 dB, GOP = 12). A \checkmark indicates that the corresponding operation or module is enabled.}
  \renewcommand{\arraystretch}{1.5}
  \footnotesize
  \begin{tabular}{>{\raggedright\arraybackslash}m{2.4cm}
                  >{\centering\arraybackslash}m{0.75cm}
                  >{\centering\arraybackslash}m{0.75cm}
                  >{\centering\arraybackslash}m{0.75cm}
                  >{\centering\arraybackslash}m{0.75cm}
                  >{\centering\arraybackslash}m{0.75cm}}
    \hline
    & $M_a$ & $M_b$ & $M_c$ & $M_d$ & $M_e$ \\
    \hline
    Rate set distillation          & \checkmark & \checkmark & \checkmark &            & \checkmark \\
    Temporal-domain feature coding &            & \checkmark &            & \checkmark & \checkmark \\
    Spatial-domain feature coding  &            &            & \checkmark & \checkmark & \checkmark \\
    \hline
    BD-CBR (\%)   & 54.6  & 34.9  & 35.5  & 4.2   & 0.0 \\  
    MACs (G)      & 729.3 & 784.1 & 742.9 & 797.7 & 797.7 \\
    Params (M)    & 30.4  & 33.3  & 32.5  & 37.2  & 35.4 \\
    Tx time (ms)  & 25.4  & 42.0  & 29.9  & 46.8  & 45.3 \\
    Rx time (ms)  & 16.8  & 25.4  & 19.3  & 27.6  & 27.0 \\
    \hline
  \end{tabular}
  \label{tab:ablation_bd}
\end{table}

\begin{table}[t]
  \centering
  \caption{Loss-term ablation on the UVG dataset (1080p) over an AWGN channel (SNR = 10 dB, GOP = 4). BD-CBR (\%) is computed with respect to VTM-17.0 + 5G LDPC. Smaller values indicate larger bandwidth savings. A \checkmark indicates that the corresponding loss term is enabled.}
  \label{tab:loss_ablation_psnr_lpips_dists}
  \renewcommand{\arraystretch}{1.25}
  \setlength{\tabcolsep}{3.4pt}
  \footnotesize
  \begin{tabular}{@{}lccc|ccc@{}}
    \toprule
    \multirow{2}{*}{\textbf{Method}} &
    \multicolumn{3}{c|}{\textbf{Loss Terms}} &
    \multicolumn{3}{c}{\textbf{BD-CBR (\%)}} \\
    \cmidrule(lr){2-4} \cmidrule(l){5-7}
    &
    $\mathcal{L}_{\mathrm{rec}}$ &
    $\mathcal{L}_{\mathrm{per}}$ &
    $\mathcal{L}_{\mathrm{adv}}$ &
    \textbf{PSNR} &
    \textbf{LPIPS} &
    \textbf{DISTS} \\
    \midrule
    VTM-17.0 + 5G LDPC
      & -- & -- & -- & 0.0 & 0.0 & 0.0 \\
    GLC-Video + 5G LDPC
      & -- & -- & -- & 369.2 & -64.2 & -91.4 \\
    \midrule
    \multirow{2}{*}{DVST}
      & \checkmark &            &            & 122.0 & 31.5  & 9.8   \\
      & \checkmark & \checkmark & \checkmark & 228.3 & -50.3 & -79.5 \\
    \midrule
    \multirow{3}{*}{PVSC (Ours)}
      & \checkmark &            &            & 59.8  & 41.3  & 45.4  \\
      & \checkmark & \checkmark &            & 206.6 & -75.3 & -91.0 \\
      & \checkmark & \checkmark & \checkmark & 204.3 & \textbf{-78.1} & \textbf{-92.2} \\
    \bottomrule
  \end{tabular}
\end{table}

\section{Conclusion}
\label{sec:conclusion}
This paper presented PVSC, a perception-aware video semantic communication framework for high-quality wireless video transmission under bandwidth and latency constraints. 
PVSC jointly optimizes semantic feature extraction, spatio-temporal entropy modeling, and rate-adaptive feature coding to improve the rate-perception-distortion tradeoff with real-time efficiency. 
Extensive experiments confirm its robustness to channel variations and its promise for next-generation wireless video applications.

\bibliographystyle{IEEEtran}
\bibliography{reference}

\begin{thebibliography}{10}
\providecommand{\url}[1]{#1}
\csname url@samestyle\endcsname
\providecommand{\newblock}{\relax}
\providecommand{\bibinfo}[2]{#2}
\providecommand{\BIBentrySTDinterwordspacing}{\spaceskip=0pt\relax}
\providecommand{\BIBentryALTinterwordstretchfactor}{4}
\providecommand{\BIBentryALTinterwordspacing}{\spaceskip=\fontdimen2\font plus
\BIBentryALTinterwordstretchfactor\fontdimen3\font minus \fontdimen4\font\relax}
\providecommand{\BIBforeignlanguage}[2]{{%
\expandafter\ifx\csname l@#1\endcsname\relax
\typeout{** WARNING: IEEEtran.bst: No hyphenation pattern has been}%
\typeout{** loaded for the language `#1'. Using the pattern for}%
\typeout{** the default language instead.}%
\else
\language=\csname l@#1\endcsname
\fi
#2}}
\providecommand{\BIBdecl}{\relax}
\BIBdecl

\bibitem{pvsc-icc}
Y.~Huang and Z.~Qin, ``Rate-efficient perception-oriented generative semantic video communication,'' in \emph{Proc. IEEE Int. Conf. Commun. Workshops (ICC Workshops)}, 2026.

\bibitem{ericsson2025}
``Ericsson mobility report, june 2025,'' White Paper, Jun. 2025.

\bibitem{polyanskiy2010finite}
Y.~Polyanskiy, H.~V. Poor, and S.~Verd{\'u}, ``Channel coding rate in the finite blocklength regime,'' \emph{IEEE Trans. Inf. Theory}, vol.~56, no.~5, pp. 2307--2359, 2010.

\bibitem{durisi2016short-packets}
G.~Durisi, T.~Koch, and P.~Popovski, ``Toward massive, ultrareliable, and low-latency wireless communication with short packets,'' \emph{Proc. IEEE}, vol. 104, no.~9, pp. 1711--1726, 2016.

\bibitem{shannon1948}
C.~E. Shannon, ``A mathematical theory of communication,'' \emph{Bell Syst. Tech. J.}, vol.~27, no.~3, pp. 379--423, 1948.

\bibitem{deepwive}
T.-Y. Tung and D.~G{\"u}nd{\"u}z, ``Deepwive: Deep-learning-aided wireless video transmission,'' \emph{IEEE J. Select. Areas Commun.}, vol.~40, no.~9, pp. 2570--2583, 2022.

\bibitem{mismatch}
Z.~Wang, A.~C. Bovik, H.~R. Sheikh, and E.~P. Simoncelli, ``Image quality assessment: From error visibility to structural similarity,'' \emph{IEEE Trans. Image Process.}, vol.~13, no.~4, pp. 600--612, 2004.

\bibitem{blau2018pd}
Y.~Blau and T.~Michaeli, ``The perception-distortion tradeoff,'' in \emph{Proc. IEEE Conf. Comput. Vis. Pattern Recog. (CVPR)}, Jun. 2018.

\bibitem{vvc}
B.~Bross, Y.-K. Wang, Y.~Ye, S.~Liu, J.~Chen, G.~J. Sullivan, and J.-R. Ohm, ``Overview of the versatile video coding ({VVC}) standard and its applications,'' \emph{IEEE Trans. Circuit Syst. Video Technol.}, vol.~31, no.~10, pp. 3736--3764, 2021.

\bibitem{dcvc}
J.~Li, B.~Li, and Y.~Lu, ``Deep contextual video compression,'' in \emph{Proc. Adv. Neural Inf. Process. Syst. (NeurIPS)}, vol.~34, 2021, pp. 18\,114--18\,125.

\bibitem{dcvc-dc}
------, ``Neural video compression with diverse contexts,'' in \emph{Proc. IEEE Conf. Comput. Vis. Pattern Recog. (CVPR)}, 2023, pp. 22\,616--22\,626.

\bibitem{dcvc-fm}
------, ``Neural video compression with feature modulation,'' in \emph{Proc. IEEE Conf. Comput. Vis. Pattern Recog. (CVPR)}, Jun. 2024, pp. 26\,099--26\,108.

\bibitem{dcvc-rt}
Z.~Jia, B.~Li, J.~Li, W.~Xie, L.~Qi, H.~Li, and Y.~Lu, ``Towards practical real-time neural video compression,'' in \emph{Proc. IEEE Conf. Comput. Vis. Pattern Recog. (CVPR)}, Nashville, TN, USA, Jun. 2025, pp. 11--25.

\bibitem{lpips}
R.~Zhang, P.~Isola, A.~A. Efros, E.~Shechtman, and O.~Wang, ``The unreasonable effectiveness of deep features as a perceptual metric,'' in \emph{Proc. IEEE Conf. Comput. Vis. Pattern Recog. (CVPR)}, Jun. 2018.

\bibitem{dists}
K.~Ding, K.~Ma, S.~Wang, and E.~P. Simoncelli, ``Image quality assessment: Unifying structure and texture similarity,'' \emph{IEEE Trans. Pattern Anal. Mach. Intell.}, vol.~44, no.~5, pp. 2567--2581, 2022.

\bibitem{plvc}
R.~Yang, R.~Timofte, and L.~Van~Gool, ``Perceptual learned video compression with recurrent conditional {GAN},'' in \emph{Proc. Int. Joint Conf. Artif. Intell. (IJCAI)}, Jul. 2022, pp. 1537--1544.

\bibitem{glc-video}
L.~Qi, Z.~Jia, J.~Li, B.~Li, H.~Li, and Y.~Lu, ``Generative latent coding for ultra-low bitrate image and video compression,'' \emph{IEEE Trans. Circuit Syst. Video Technol.}, vol.~35, no.~10, pp. 10\,500--10\,515, 2025.

\bibitem{10639525}
Z.~Qin, L.~Liang, Z.~Wang, S.~Jin, X.~Tao, W.~Tong, and G.~Y. Li, ``{AI} empowered wireless communications: From bits to semantics,'' \emph{Proc. IEEE}, vol. 112, no.~7, pp. 621--652, Jul. 2024.

\bibitem{Task-Oriented}
H.~Xie, Z.~Qin, X.~Tao, and K.~B. Letaief, ``Task-oriented multi-user semantic communications,'' \emph{IEEE J. Select. Areas Commun.}, vol.~40, no.~9, pp. 2584--2597, Sept. 2022.

\bibitem{deepsc}
H.~Xie, Z.~Qin, G.~Y. Li, and B.-H. Juang, ``Deep learning enabled semantic communication systems,'' \emph{IEEE Trans. Signal Process.}, vol.~69, pp. 2663--2675, Apr. 2021.

\bibitem{robustsc-speech}
Z.~Weng, Z.~Qin, and G.~Y. Li, ``Robust semantic communications for speech transmission,'' in \emph{Proc. IEEE Int. Conf. Acoust. Speech Signal Process. (ICASSP)}, 2025, pp. 1--5.

\bibitem{ntscc}
J.~Dai, S.~Wang, K.~Tan, Z.~Si, X.~Qin, K.~Niu, and P.~Zhang, ``Nonlinear transform source-channel coding for semantic communications,'' \emph{IEEE J. Select. Areas Commun.}, vol.~40, no.~8, pp. 2300--2316, Aug. 2022.

\bibitem{plit}
G.~Zhang, H.~Li, Y.~Cai, Q.~Hu, G.~Yu, and Z.~Qin, ``Progressive learned image transmission for semantic communication using hierarchical vae,'' \emph{IEEE Trans. Cognit. Comm. Netw.}, vol.~11, no.~6, pp. 3640--3654, 2025.

\bibitem{Quaddeepsc}
Y.~Huang and Z.~Qin, ``Image semantic communication with quadtree partition-based coding,'' \emph{IEEE J. Select. Areas Commun.}, vol.~44, pp. 2765--2778, 2026.

\bibitem{yjk2025}
J.~Ying, Z.~Qin, Y.~Feng, L.~Wang, and X.~Tao, ``Joint semantic-channel coding and modulation for token communications,'' \emph{IEEE Trans. Wirel. Commun.}, vol.~25, pp. 8179--8193, 2026.

\bibitem{dvst}
S.~Wang, J.~Dai, Z.~Liang, K.~Niu, Z.~Si, C.~Dong, X.~Qin, and P.~Zhang, ``Wireless deep video semantic transmission,'' \emph{IEEE J. Select. Areas Commun.}, vol.~41, no.~1, pp. 214--229, 2023.

\bibitem{DVSC-gcwkp}
H.~Niu, L.~Wang, Z.~Lu, K.~Du, and X.~Wen, ``Deep learning enabled video semantic transmission against multi-dimensional noise,'' in \emph{2023 IEEE Globecom Workshops (GC Wkshps)}, 2023, pp. 1267--1272.

\bibitem{wvjscc}
Y.~Huang and Z.~Qin, ``Wireless video transmission with joint semantic-channel coding,'' in \emph{Proc. IEEE Globecom Workshops (GC Wkshps)}, 2024, pp. 1--6.

\bibitem{mdvsc}
Z.~Bao, H.~Liang, C.~Dong, C.~Li, X.~Xu, and P.~Zhang, ``Mdvsc—efficient wireless model division video semantic communication,'' \emph{IEEE Internet Things J.}, vol.~12, no.~2, pp. 1109--1124, 2025.

\bibitem{URLLC-iccwkp}
C.~Liang, X.~Deng, Y.~Sun, R.~Cheng, L.~Xia, D.~Niyato, and M.~A. Imran, ``Vista: Video transmission over a semantic communication approach,'' in \emph{Proc. IEEE Int. Conf. Commun. Workshops (ICC Workshops)}, 2023, pp. 1777--1782.

\bibitem{SemCom-video}
Z.~Zhang, Q.~Yang, S.~He, and Z.~Shi, ``Bidirectional motion-enhanced semantic communication for wireless video transmission,'' \emph{IEEE Internet Things J.}, vol.~13, no.~8, pp. 15\,607--15\,620, 2026.

\bibitem{PSD-GSC}
N.~Li, Y.~Deng, and D.~Niyato, ``Goal-oriented semantic communication for wireless video transmission via generative ai,'' \emph{IEEE Trans. Wirel. Commun.}, vol.~25, pp. 10\,841--10\,854, 2026.

\bibitem{oar-vsc}
Q.~Du, Y.~Duan, Q.~Yang, X.~Tao, and M.~Debbah, ``Object-attribute-relation representation-based video semantic communication,'' \emph{IEEE J. Select. Areas Commun.}, vol.~43, no.~7, pp. 2446--2461, 2025.

\bibitem{wsc-vc}
P.~Jiang, C.-K. Wen, S.~Jin, and G.~Y. Li, ``Wireless semantic communications for video conferencing,'' \emph{IEEE J. Select. Areas Commun.}, vol.~41, no.~1, pp. 230--244, 2023.

\bibitem{syncsc}
Y.~Tian, J.~Ying, Z.~Qin, Y.~Jin, and X.~Tao, ``Synchronous multi-modal semantic communication system with packet-level coding,'' \emph{IEEE Trans. Wirel. Commun.}, vol.~24, no.~5, pp. 3684--3697, 2025.

\bibitem{10093128}
L.~Agnolucci, L.~Galteri, M.~Bertini, and A.~D. Bimbo, \emph{IEEE Trans. Multimedia}.

\bibitem{videoqa-sc}
J.~Guo, W.~Chen, Y.~Sun, J.~Xu, and B.~Ai, ``{VideoQA-SC}: Adaptive semantic communication for video question answering,'' \emph{IEEE J. Select. Areas Commun.}, vol.~43, no.~7, pp. 2462--2477, 2025.

\bibitem{bjornson2017massive}
E.~Bj{\"o}rnson, J.~Hoydis, and L.~Sanguinetti, ``Massive {MIMO} networks: Spectral, energy, and hardware efficiency,'' \emph{Found. Trends Signal Process.}, vol.~11, no. 3-4, pp. 154--655, 2017.

\bibitem{deepjscc}
E.~Bourtsoulatze, D.~B. Kurka, and D.~G{\"u}nd{\"u}z, ``Deep joint source-channel coding for wireless image transmission,'' \emph{IEEE Trans. Cognit. Comm. Netw.}, vol.~5, no.~3, pp. 567--579, May. 2019.

\bibitem{balle2016end}
J.~Ball{\'e}, V.~Laparra, and E.~P. Simoncelli, ``End-to-end optimized image compression,'' in \emph{Proc. Int. Conf. Learn. Represent. (ICLR)}, Toulon, France, Apr. 2017.

\bibitem{png-guide}
G.~Roelofs and R.~Koman, \emph{{PNG}: The Definitive Guide}.\hskip 1em plus 0.5em minus 0.4em\relax USA: O'Reilly \& Associates, Inc., 1999.

\bibitem{taming-transformers}
P.~Esser, R.~Rombach, and B.~Ommer, ``Taming transformers for high-resolution image synthesis,'' in \emph{Proc. IEEE Conf. Comput. Vis. Pattern Recog. (CVPR)}, Jun. 2021, pp. 12\,873--12\,883.

\bibitem{gan}
I.~J. Goodfellow, J.~Pouget-Abadie, M.~Mirza, B.~Xu, D.~Warde-Farley, S.~Ozair, A.~Courville, and Y.~Bengio, ``Generative adversarial nets,'' in \emph{Proc. Adv. Neural Inf. Process. Syst. (NeurIPS)}, vol.~27, 2014.

\bibitem{he2021checkerboard}
D.~He, Y.~Zheng, B.~Sun, Y.~Wang, and H.~Qin, ``Checkerboard context model for efficient learned image compression,'' in \emph{Proc. IEEE Conf. Comput. Vis. Pattern Recog. (CVPR)}, 2021, pp. 14\,771--14\,780.

\bibitem{hu2023robust}
Q.~Hu, G.~Zhang, Z.~Qin, Y.~Cai, G.~Yu, and G.~Y. Li, ``Robust semantic communications with masked {VQ-VAE} enabled codebook,'' \emph{IEEE Trans. Wirel. Commun.}, vol.~22, no.~12, pp. 8707--8722, Dec. 2023.

\bibitem{vimeo}
T.~Xue, B.~Chen, J.~Wu, D.~Wei, and W.~T. Freeman, ``Video enhancement with task-oriented flow,'' \emph{Int. J. Comput. Vis.}, vol. 127, no.~8, pp. 1106--1125, 2019.

\bibitem{bvi-dvc}
D.~Ma, F.~Zhang, and D.~R. Bull, ``{BVI-DVC}: A training database for deep video compression,'' \emph{IEEE Trans. Multimedia}, vol.~24, pp. 3847--3858, 2021.

\bibitem{vgg}
K.~Simonyan and A.~Zisserman, ``Very deep convolutional networks for large-scale image recognition,'' \emph{arXiv preprint arXiv:1409.1556}, 2014.

\bibitem{bjontegaard2001calculation}
G.~Bjontegaard, ``Calculation of average psnr differences between rd-curves,'' \emph{ITU SG16 Doc. VCEG-M33}, 2001.

\bibitem{hevcclass}
D.~Flynn, K.~Sharman, and C.~Rosewarne, ``Common test conditions and software reference configurations for {HEVC} range extensions,'' \emph{JCT-VC Doc. JCTVC-N1006}, vol.~16, p.~6, 2013.

\bibitem{uvg}
A.~Mercat, M.~Viitanen, and J.~Vanne, ``{UVG} dataset: 50/120fps 4k sequences for video codec analysis and development,'' in \emph{Proc. ACM Multimedia Syst. Conf. (MMSys)}, Istanbul, Turkey, 2020, pp. 297--302.

\bibitem{mcl-jcv}
H.~Wang, W.~Gan, S.~Hu, J.~Y. Lin, L.~Jin, L.~Song, P.~Wang, I.~Katsavounidis, A.~Aaron, and C.-C.~J. Kuo, ``{MCL-JCV}: A {JND}-based {H.264/AVC} video quality assessment dataset,'' in \emph{Proc. IEEE Int. Conf. Image Process. (ICIP)}, 2016, pp. 1509--1513.

\bibitem{FFmpeg}
``{FFmpeg} reference software,'' \url{https://www.ffmpeg.org/}, accessed: 2025-04-13.

\bibitem{hevc_hm}
``{HEVC} official test model,'' \url{https://hevc.hhi.fraunhofer.de}, accessed: 2025-04-13.

\bibitem{vvc_vtm}
``{VVC} official test model,'' \url{https://vcgit.hhi.fraunhofer.de/jvet/VVCSoftware_VTM}, accessed: 2025-04-13.

\bibitem{richardson2018design}
T.~Richardson and S.~Kudekar, ``Design of low-density parity check codes for 5g new radio,'' \emph{IEEE Commun. Mag.}, vol.~56, no.~3, pp. 28--34, Mar. 2018.

\bibitem{3gpp38214}
{3GPP}, ``{3GPP TS 38.214 version 16.2.0 Release 16: 5G; NR; Physical layer procedures for data},'' \url{https://www.etsi.org/deliver/etsi_ts/138200_138299/138214/16.02.00_60/ts_138214v160200p.pdf}, 2020, accessed: 2025-04-13.

\bibitem{sionna}
J.~Hoydis, S.~Cammerer, F.~{Ait Aoudia}, M.~Nimier-David, L.~Maggi, G.~Marcus, A.~Vem, and A.~Keller, ``Sionna,'' 2022, https://nvlabs.github.io/sionna/.

\end{thebibliography}

\end{document}